\documentclass{PoS}

 \usepackage{graphicx,amsmath,amssymb}

\title{AdS/QCD,  Light-Front Holography, and Sublimated Gluons}
\ShortTitle{AdS/QCD,  Light-Front Holography, and Sublimated Gluons}
\author{\speaker{Stanley J. Brodsky}\\
              SLAC National Accelerator Laboratory, \\
         Stanford University, Stanford, CA 94309, USA \\
          E-mail: \email{sjbth@slac.stanford.edu}}
\author{Guy F. de T\'eramond\\
       Universidad de Costa Rica,\\
San Jos\'e, Costa Rica\\
E-mail:\email{gdt@asterix.crnet.cr}}

\abstract{Gauge/gravity duality leads to a simple, analytical, and phenomenologically compelling nonperturbative approximation to the full light-front QCD Hamiltonian.  This approach, called  ``Light-Front Holography'', successfully describes the spectroscopy of light-quark meson and baryons, their elastic and transition form factors, and other hadronic properties. 
The bound-state
Schr\"odinger and Dirac equations of the soft-wall AdS/QCD model predict linear Regge trajectories which have the same slope in orbital angular momentum $L$ and radial
quantum number $n$ for both mesons and baryons.  
Light-front holography connects the fifth-dimensional coordinate of AdS space $z$ to an invariant impact separation variable $\zeta$ in $3+1$ space at fixed light-front time. A key feature is the determination of 
the frame-independent light-front wavefunctions of hadrons --  the relativistic 
analogs of the Schr\"odinger wavefunctions of atomic physics  which allow one to compute form
factors, transversity distributions, spin properties of  the valence quarks,  jet hadronization, and other hadronic observables.
One thus obtains a one-parameter color-confining model for hadron physics at the amplitude level.
AdS/QCD also predicts the form of the non-perturbative effective coupling $\alpha_s^{AdS}(Q)$ and its $\beta$-function
with an infrared fixed point which agrees with the effective
coupling $\alpha_{g_1}(Q^2)$ extracted from  measurements of the Bjorken sum rule below $Q^2  \! < 1 ~ {\rm GeV}^2$.
This is consistent with a flux-tube interpretation of QCD where soft gluons with  virtualities $Q^2  \! < 1 ~ {\rm GeV}^2$ are sublimated into a color-confining potential for quarks.  We discuss a number of phenomenological hadronic properties which support this picture.}

\FullConference{International Workshop on QCD Green's Functions, Confinement and Phenomenology\\
                 5-9 September 2011\\
                 Trento, Italy}

\begin{document}

\section{Introduction}

A central problem of hadron physics is how to compute the properties and spectroscopy of hadrons in terms of the confined quark and gluon degrees of freedom of the QCD Lagrangian.
As we shall discuss, 
AdS/QCD, which uses  the 
AdS/CFT correspondence~\cite{Maldacena:1997re}
between higher-dimensional anti--de Sitter (AdS) space and conformal field theories in physical space-time as a useful guide,
combined with ``Light-Front Holography'',~\cite{deTeramond:2008ht} leads to an analytic and 
phenomenologically compelling first approximation to QCD.
Light-front  holographic methods  were originally
introduced~\cite{Brodsky:2006uqa, Brodsky:2007hb} by matching the  electromagnetic (EM) 
current matrix elements in AdS space~\cite{Polchinski:2002jw} with
the corresponding expression  using light-front  quantization in
physical space time.~\cite{Drell:1969km, West:1970av}  It has also been shown that one obtains  identical holographic mapping using the matrix elements of the
energy-momentum tensor~\cite{Brodsky:2008pf} by perturbing the AdS
metric  around its static solution.~\cite{Abidin:2008ku}

Light front (LF) holographic methods
allow one to project the functional dependence of the wavefunction $\Phi(z)$ computed  in the  AdS fifth dimension to the  hadronic frame-independent light-front wavefunction $\psi(x_i, \vec b_{\perp i})$ in $3+1$ physical space-time. The variable $z $ maps  to a
LF variable $ \zeta(x_i, \vec b_{\perp i})$ which measures the invariant separation of the constituents within a hadron at equal light-front time $\tau = t + z/c = x^+ = x^0 + x^3$.
This leads to single-variable light-front Schr\"odinger and Dirac equations whose eigensolutions  determine the spectrum of hadrons for general spin and orbital angular momentum.  The resulting light-front wavefunctions allow one to compute form
factors, transversity distributions, spin properties of  the valence quarks,  jet hadronization, and other hadronic observables.
One thus obtains a one-parameter color-confining model for hadron physics at the amplitude level.

In atomic physics, one computes the properties of electrons within an atom from its Schr\"odinger wavefunction, the eigenfunction of the nonrelativistic approximation  to the exact QED Hamiltonian.  In principle, one could also  calculate hadronic spectroscopy and wavefunctions by solving for the eigenstates of the QCD Hamiltonian: 
$H \vert  \Psi \rangle = E \vert \Psi \rangle$
 at fixed time $t.$ However, this traditional method -- called the ``instant form'' by Dirac,~\cite{Dirac:1949cp} is  plagued by complex vacuum and relativistic effects, as well 
as  by  the fact that the boost of such fixed-$t$ wavefunctions away from the hadron's rest frame is an intractable dynamical problem.  
In contrast,  quantization at fixed light-front time $\tau$,  the ``front-form'' of Dirac~\cite{Dirac:1949cp}, is a powerful non-perturbative alternative, providing a boost-invariant framework for describing the structure of hadrons in terms of their 
constituents.
The simple structure of the light-front  vacuum allows an unambiguous
definition of the partonic content of a hadron in QCD in terms of the hadronic light-front wavefunctions (LFWFs),
the underlying link between large distance hadronic states and the
constituent degrees of freedom at short distances. Measurements of the constituents of a bound state such as in form factors or in deep inelastic scattering are determined at the same light-front time $\tau$, along the front of a light-wave as in a flash picture.  In contrast, the constituents of a bound state in an instant-form wavefunction must be measured at the same instant time $t$,  requiring the exact synchrony in time of many simultaneous probes. 
Calculating a hadronic form factor in the traditional instant form requires boosting the hadron's wavefunction from the initial to final state, a dynamical problem  as difficult as solving QCD itself. The boost invariance of  LFWFs contrasts dramatically with the complexity of  boosting the wavefunctions defined at fixed time $t.$~\cite{Brodsky:1968ea}  
Worse, evaluating a current matrix element in the instant form
requires
computing the interaction of the probe with all of connected currents fluctuating in the acausal instant form vacuum, and  each contributing diagram is frame-dependent.

Light-front quantization is thus the ideal framework to describe the
structure of hadrons in terms of their quark and gluon degrees of freedom.  The
constituent spin and orbital angular momentum properties of the
hadrons are also encoded in the LFWFs.  
The total  angular momentum projection~\cite{Brodsky:2000ii} 
$J^z = \sum_{i=1}^n  S^z_i + \sum_{i=1}^{n-1} L^z_i$ 
is conserved, Fock-state by Fock-state, and by every interaction in the LF Hamiltonian.
The empirical observation that the quark spin $S^z$ contributes only a small fraction of the nucleon's total angular momentum $J^z$ highlights the importance of quark orbital angular momentum.  In fact, the nucleon anomalous moment, the Pauli form factor, and single-spin asymmetries, such as the Sivers effect, would vanish unless the quarks carry nonzero $L^z$.

\section{Light-Front Quantization and Light-Front Wavefunctions}

In the light-front formalism, one  sets boundary conditions at fixed $\tau$ and then evolves the system using the LF Hamiltonian
$P^-  \! =  \! P^0 \! - P^3 = i {d/d \tau}$.  The invariant Hamiltonian $H_{LF}  \! =  P_\mu P^\mu = P^+ P^- \! - \vec P^2_\perp$
has eigenvalues $\mathcal{M}^2$ where $\mathcal{M}$
is the physical hadron mass.
The Heisenberg equation for QCD on the light-front thus takes the form 
\begin{equation} \label{LFH}
H_{LF} \vert \Psi_H \rangle = \mathcal{M}^2_H \vert \Psi_H \rangle ,
\end{equation}
 where $H_{LF}$ is determined canonically from the QCD Lagrangian.~\cite{Brodsky:1997de} 
Its eigenfunctions are the light-front eigenstates which define the frame-independent light-front wavefunctions, and
its eigenvalues yield the hadronic spectrum, the bound states as well as the continuum.  The  projection of the eigensolutions on the free Fock basis gives the $n$-parton LF wavefunctions
$\psi_{n/H} = \langle n \vert \Psi_H \rangle$ needed for phenomenology.   Heisenberg's problem on the light-front can be solved numerically using discretized light-front quantization (DLCQ)~\cite{Pauli:1985ps} by applying anti-periodic boundary conditions in $\sigma = x^0-x^3.$  This method has been used successfully to solve many lower dimension quantum field theories.~\cite{Brodsky:1997de}

The light-front Fock state wavefunctions
$\psi_{n/H}(x_i, \vec k_\perp, \lambda_i)$ are functions of LF momentum fractions $x_i = {k^+_i\over P^+} = {k^0_i+k^3_i\over P^0+P^3}$ with $\sum^n_{i=1} x_i =1,$  relative transverse momenta satisfying 
$\sum^n_{i=1} \vec k_{\perp i} = 0$,
 and spin projections $\lambda_i.$  Remarkably, the LFWFs are independent of the hadron's total momentum $P^+ = P^0 \! + P^3$, so that once they are known in one frame, they are known in all frames; Wigner transformations and Melosh rotations
are not required. 
Light-cone gauge $A^+=0$ is particularly useful since there are no ghosts
and one has a direct physical interpretation of  orbital angular
momentum.  They also allows one to formulate hadronization in inclusive and exclusive reactions at the amplitude level.

In general, hadron eigenstates 
have components with different orbital angular momentum; for example,  the proton eigenstate in LF holographic QCD  with massless quarks has $L=0$ and $L=1$ light-front Fock components with equal probability.   Higher Fock states with extra quark-anti quark pairs also arise.   The resulting LFWFs then lead to a new range of hadron phenomenology,  including the possibility to compute the hadronization of quark and gluon jets at the amplitude level. The soft-wall model also predicts the form of the non-perturbative effective coupling and its $\beta$-function.~\cite{Brodsky:2010ur}

A key example of the utility of the light-front is the Drell-Yan West formula~\cite{Drell:1969km, West:1970av}
for the spacelike form factors of electromagnetic currents, given as simple convolutions of initial and final LFWFs. The gauge-invariant distribution
amplitude $\phi_H(x_i,Q)$, defined from the integral over the
transverse momenta ${\vec k}^2_{\perp i} \le Q^2$ of the valence
(smallest $n$) Fock state, 
provides a fundamental measure of the
hadron at the amplitude level;~\cite{Lepage:1979zb, Efremov:1979qk}
they  are the nonperturbative inputs to the factorized form of hard
exclusive amplitudes and exclusive heavy hadron decays in
pQCD. At high momentum where one can iterate the hard scattering kernel, this yields dimensional counting rules, factorization theorems, and ERBL evolution of the distribution amplitudes.

Given the frame-independent light-front
wavefunctions $\psi_{n/H}(x_i, \vec k_{\perp i}, \lambda_i )$, one can compute virtually all exclusive and inclusive hadron observables. 
For example, the valence and sea quark and gluon
distributions which are measured in deep inelastic lepton scattering (DIS)
are defined from the squares of the LFWFs summed over all Fock
states $n$.  Form factors,
exclusive weak transition
amplitudes~\cite{Brodsky:1998hn} such as $B\to \ell \nu \pi$,  and
the generalized parton distributions~\cite{Brodsky:2000xy} measured
in deeply virtual Compton scattering  $\gamma^* p \to \gamma p$ are (assuming the ``handbag"
approximation) overlaps of the initial and final LFWFs with $n
=n^\prime$ and $n =n^\prime+2$. The resulting distributions obey the DGLAP and
ERBL evolution equations as a function of the maximal invariant
mass, thus providing a physical factorization
scheme.~\cite{Lepage:1980fj} In each case, the derived quantities
satisfy the appropriate operator product expansions, sum rules, and
evolution equations. At large $x$ where the struck quark is
far-off shell, DGLAP evolution is quenched,~\cite{Brodsky:1979qm} so
that the fall-off of the DIS cross sections in $Q^2$ satisfies Bloom-Gilman
inclusive-exclusive duality at fixed $W^2.$
ERBL evolution of the distribution amplitudes and DGLAP evolution of structure functions are automatically satisfied. 
Transversity observables can also be computed from the LFWFs; 
however, in the case of pseudo-T-odd observables, one must include the lensing effect of final state or initial state interactions.

A fundamental theorem for gravity can be derived from the equivalence principle:  the anomalous gravitomagnetic moment defined from the spin-flip  matrix element of the energy-momentum tensor is identically zero $B(0)=0$.~\cite{Teryaev:1999su} This theorem can be proven in  the light-front formalism Fock state by Fock state.~\cite{Brodsky:2000ii} The LF vacuum is trivial up to zero modes in the front form, thus eliminating contributions to the cosmological constant from QED or QCD.~\cite{Brodsky:2009zd}

One can solve the LF Hamiltonian problem for theories  in one-space and one-time  by Heisenberg matrix diagonalization. For example, the complete set of discrete and continuum eigensolutions of mesons and baryons  in QCD(1+1) can be obtained to any desired precision for general color,  multiple flavors, and general quark masses using the discretized light-cone quantized (DLCQ) method.~\cite{Pauli:1985ps,Hornbostel:1988fb}  The  DLCQ approach can in principle be applied to QED(3+1) and QCD(3+1). This nonperturbative method has no fermion doubling, is in Minkowski space, and frame-independent; however,  in practice, the resulting large matrix diagonalization problem has found to be computational challenging, so alternative methods and approximations such as those based on AdS/QCD are necessary.

\section{AdS/QCD}

One of the most significant theoretical advances in recent years has
been the application of the AdS/CFT
correspondence~\cite{Maldacena:1997re} between string theories
defined in 5-dimensional AdS space and
conformal field theories in physical space-time,
to study  the dynamics of strongly coupled quantum field theories.
The essential principle
underlying the AdS/CFT approach to conformal gauge theories is the
isomorphism of the group of Poincar\'{e} and conformal transformations
$SO(4,2)$ to the group of isometries of Anti-de Sitter space.  The
AdS metric is
\begin{equation} \label{eq:AdSz}
ds^2 = \frac{R^2}{z^2}(\eta_{\mu \nu} dx^\mu
 dx^\nu - dz^2),
 \end{equation}
which is invariant under scale changes of the
coordinate in the fifth dimension $z \to \lambda z$ and $ x^\mu \to
\lambda x^\mu$.  Thus one can match scale transformations of the
theory in $3+1$ physical space-time to scale transformations in the
fifth dimension $z$.
In the AdS/CFT duality, the amplitude $\Phi(z)$ represents the
extension of the hadron into the additional fifth dimension.  The
behavior of
$\Phi(z) \to z^\tau$ at $z \to 0$
matches the twist-dimension $\tau$ of the hadron at short distances.

QCD is not conformal but  there is in fact much empirical evidence from lattice gauge theory,~\cite{Furui:2006py} Dyson Schwinger theory,~\cite{vonSmekal:1997is}  and empirical effective charges,~\cite{Deur:2005cf} that the QCD $\beta$-function vanishes in the infrared.~\cite{Deur:2008rf}  The QCD infrared fixed point arises since the propagators of the confined quarks and gluons in the  loop integrals contributing to the $\beta$-function have a maximal wavelength.~\cite{Brodsky:2008be} The decoupling of quantum loops in the infrared is analogous to QED where vacuum polarization corrections to the photon propagator decouple at $Q^2 \to 0$.
Since there is a  window where the QCD coupling is
large and approximately constant,
QCD resembles a conformal theory for massless quarks. Thus, even though QCD is not conformally invariant,
one can use the mathematical representation of the conformal group
in five-dimensional Anti-de Sitter space to construct an analytic
first approximation to the theory.

\subsection{AdS/QCD Models}

We thus begin with a conformal approximation to QCD to model an effective dual gravity description in AdS space. The  five-dimensional AdS$_5$ geometrical representation of the conformal group represents scale transformations within the conformal window.  Confinement can be introduced with a sharp cut-off in the infrared region of AdS space, as in the ``hard-wall'' model,~\cite{Polchinski:2001tt}
 or, more successfully,  using a dilaton background in the fifth dimension to produce a smooth cutoff at large distances
as  in the ``soft-wall'' model.~\cite{Karch:2006pv}
We assume a dilaton profile 
 $\exp(+\kappa^2 z^2)$~\cite{deTeramond:2009xk}
with  opposite sign  to that of Ref.~\cite{Karch:2006pv}; however as shown in Ref.~\cite{Gutsche:2011vb}, one can relate the solutions by a canonical transformation.  The positive dilaton background leads to the harmonic potential~\cite{deTeramond:2009xk} 
$U(z) = \kappa^4 z^2 + 2 \kappa^2(L+S-1)$~in the fifth dimension coordinate $z$.
The resulting spectrum reproduces linear Regge trajectories for mesons:
${\cal M}^2(S,L,n) = 4 \kappa^2(n + L + S/2)$,
where ${\cal M}^2(S,L,n) $ is proportional to the internal
spin,  orbital angular momentum $L$ and the
principal radial quantum number $n$. 

The conformal invariance can also be broken by the introduction of an additional warp factor 
in the AdS metric to include confinement forces:
\begin{equation}  \label{gE}
ds^2 = \frac{R^2}{z^2} e^{\lambda(z)} \left( \eta_{\mu \nu} dx^\mu dx^\nu - dz^2\right).
\end{equation}
The results are equivalent to introducing a dilaton profile, provided that $\varphi(z)  \to \frac{d-1}{2} \lambda(z)$, or $\varphi \to \frac{3}{2} \lambda$ for AdS$_5$. 
The modified metric  can be interpreted in AdS space as a gravitational potential
for an object of mass $m$  in the fifth dimension:
$V(z) = mc^2 \sqrt{g_{00}} = mc^2 R \, e^{\pm 3 \kappa^2 z^2/4}/z$.
In the case of the negative solution, the potential decreases monotonically, and thus an object in AdS will fall to infinitely large
values of $z$.  For the positive solution, the potential is non-monotonic and has an absolute minimum at $z_0 \sim 1/\kappa$.
Furthermore, for large values of $z$ the gravitational potential increases exponentially,  confining any object  to distances
$\langle z \rangle \sim 1/\kappa$.~\cite {deTeramond:2009xk}  We thus will choose the confining positive sign dilaton solution.~\cite{deTeramond:2009xk, Andreev:2006ct} This additional warp factor leads to a well-defined scale-dependent effective coupling.

The identification of orbital angular momentum of the constituents is a key element in the description of the internal structure of hadrons using holographic principles. In our approach  quark and gluon degrees of freedom are explicitly introduced in the gauge/gravity correspondence,~\cite{Brodsky:2003px} in contrast with the usual
AdS/QCD framework~\cite{Erlich:2005qh,DaRold:2005zs} where axial and vector currents become the primary entities as in effective chiral theory.
Unlike the top-down string theory approach,  one is not limited to hadrons of maximum spin
$J \le 2$, and one can study baryons with finite color $N_C=3.$

Glazek and Schaden~\cite{Glazek:1987ic} have shown that a  harmonic oscillator confining potential naturally arises as an effective potential between heavy quark states when one stochastically eliminates higher gluonic Fock states. Also, Hoyer~\cite{Hoyer:2009ep} has argued that the Coulomb  and   linear  potentials are uniquely allowed in the Dirac equation at the classical level. The linear potential  becomes a harmonic oscillator potential in the corresponding Klein-Gordon equation.

\section{Light-Front Holography}

Light-front  holography \cite{deTeramond:2008ht, Brodsky:2006uqa, Brodsky:2007hb} provides a remarkable connection between
the equations of motion in AdS space and
the Hamiltonian formulation of QCD in physical space-time quantized
on the light front  at fixed LF time $\tau$.
This correspondence provides a direct connection between the hadronic amplitudes $\Phi(z)$  in AdS space  with  LF wavefunctions $\phi(\zeta)$ describing the quark and gluon constituent structure of hadrons in physical space-time.
In the case of a meson, $\zeta = \sqrt{x(1-x) {\vec b}^2_\perp}$ is a Lorentz invariant coordinate which measures
the distance between the quark and antiquark; it is analogous to the radial coordinate $r$ in the Schr\"odinger equation. Here $\vec b_\perp$ is the Fourier conjugate of the transverse
momentum $\vec k_\perp$.
In effect $\zeta$ represents the off-light-front energy shell and invariant mass dependence of the bound state; it allows the separation of the dynamics of quark and gluon binding from the kinematics of constituent spin and internal orbital angular momentum.~\cite{deTeramond:2008ht}
The resulting equation for the mesonic $q \bar q$ bound states at fixed light-front time  has the form of a single-variable relativistic Lorentz invariant
 Schr\"odinger equation~\cite{deTeramond:2008ht}
\begin{equation} \label{eq:QCDLFWE}
\left(-\frac{d^2}{d\zeta^2}
- \frac{1 - 4L^2}{4\zeta^2} + U(\zeta) \right)
\phi(\zeta) = \mathcal{M}^2 \phi(\zeta),
\end{equation}
where the confining potential is $ U(\zeta) = \kappa^4 \zeta^2 + 2 \kappa^2(L+S-1)$
in the soft-wall model
with a positive-sign dilaton-modified AdS metric.~\cite{deTeramond:2009xk}
Its eigenvalues determine the hadronic spectra and its eigenfunctions are related to the light-front wavefunctions
of hadrons for general spin and orbital angular momentum. This
LF wave equation serves as a semiclassical first approximation to QCD,
and it is equivalent to the equations of motion which describe the
propagation of spin-$J$ modes in  AdS space.  The resulting light-front wavefunctions provide a fundamental description of the structure and internal dynamics of hadronic states in terms of their constituent quark and gluons.
There is only one parameter, the mass scale $\kappa \sim 1/2$ GeV, which enters the confinement potential. In the case of mesons $S=0,1$ is the combined spin of the $q $ and $ \bar q $ state, $L$ is their relative orbital angular momentum as determined by the hadronic light-front wavefunctions.

The mapping between the LF invariant variable $\zeta$ and the fifth-dimension AdS coordinate $z$ was originally obtained
by matching the expression for electromagnetic current matrix
elements in AdS space  with the corresponding expression for the
current matrix element, using LF  theory in physical space
time.~\cite{Brodsky:2006uqa}   It has also been shown that one
obtains the identical holographic mapping using the matrix elements
of the energy-momentum tensor,~\cite{Brodsky:2008pf,Abidin:2008ku} thus  verifying  the  consistency of the holographic
mapping from AdS to physical observables defined on the light front.

\section{The Hadron Spectrum
in Light-Front AdS/QCD}

\begin{figure}[h]
\begin{center}
\includegraphics[width=6.8cm]{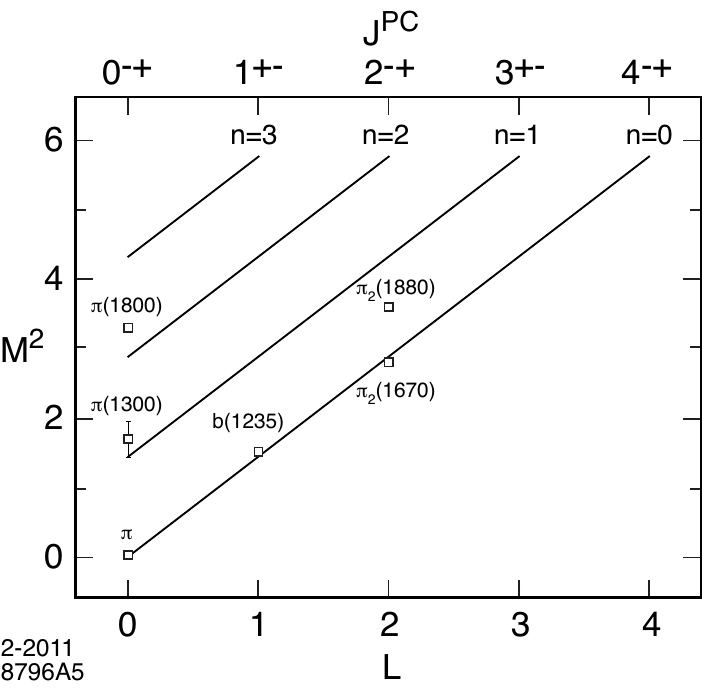}  ~~~
\includegraphics[width=6.8cm]{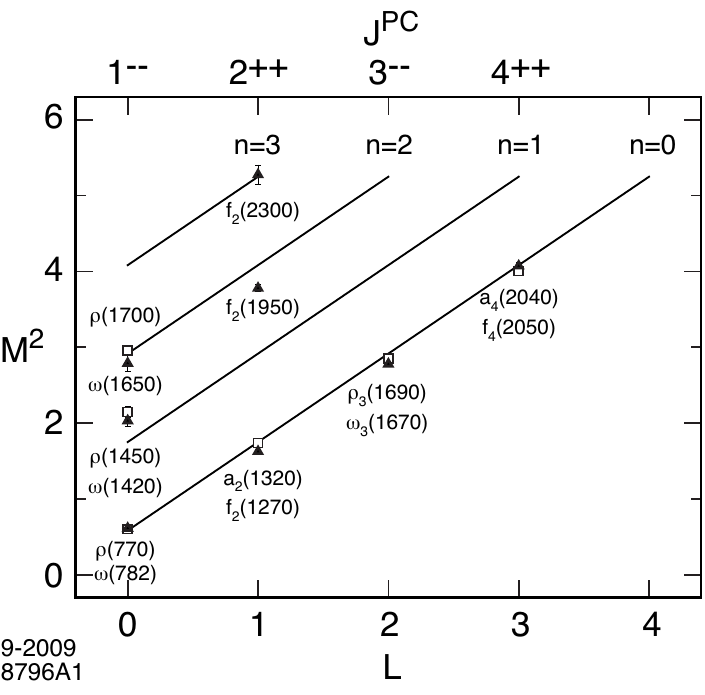}
 \caption{Parent and daughter Regge trajectories for (a) the $\pi$-meson family with
$\kappa= 0.6$ GeV; and (b) the  $I\!=\!1$ $\rho$-meson
 and $I\!=\!0$  $\omega$-meson families with $\kappa= 0.54$ GeV.}
\label{pionVM}
\end{center}
\end{figure}

The meson spectrum predicted by  Eq. \ref{eq:QCDLFWE} has a string-theory Regge form
${\cal M}^2 = 4 \kappa^2(n+ L+S/2)$; {\it i.e.}, the square of the eigenmasses are linear in both the orbital angular momentum $L$ and $n$, where $n$ counts the number of nodes  of the wavefunction in the radial variable $\zeta$.
The spectrum also depends on the internal spin S.
This is illustrated for the pseudoscalar and vector meson spectra in Fig. \ref{pionVM},
where the data are from Ref.~\cite{FERMILAB-PUB-10-665-PPD}.
The pion ($S=0, n=0, L=0$) is massless for zero quark mass, consistent with the chiral invariance of massless
quarks in
QCD.  Thus one can compute the hadron spectrum by simply adding  $4 \kappa^2$ for a unit change in the radial quantum number, $4 \kappa^2$ for a change in one unit in the orbital quantum number  $L$ and $2 \kappa^2$ for a change of one unit of spin $S$. Remarkably, the same rule holds for three-quark baryons as we shall show below.

\begin{figure}[h]
\begin{center}
 \includegraphics[width=8.0cm]{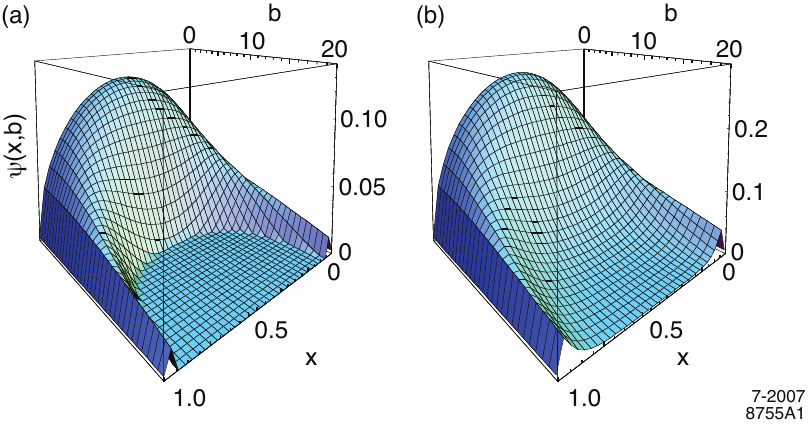}
 \caption{ Pion LF wavefunction $\psi_\pi(x, \vec{b}_\perp$) for the  AdS/QCD (a) hard-wall  and (b) soft-wall   models.  }
 \label{LFWFPion}
\end{center}
\end{figure}

The eigensolutions of  Eq. \ref{eq:QCDLFWE} provide the light-front wavefunctions of the valence Fock state of the hadrons $\psi(x,
\vec b_\perp)$~\cite{Brodsky:2006uqa}  as illustrated for the pion in Fig. \ref{LFWFPion} for the soft-wall (a) and hard-wall (b) models.   The resulting distribution amplitude has a
broad form $\phi_\pi(x) \sim \sqrt{x(1-x)}$ which is compatible with moments determined from lattice gauge theory. One can then immediately
compute observables such as hadronic form factors (overlaps of LFWFs), structure functions (squares of LFWFs), as well as the generalized parton
distributions and distribution amplitudes which underly hard exclusive reactions. For example, hadronic form factors can be predicted from the
overlap of LFWFs in the Drell-Yan West formula.

Individual hadrons in AdS/QCD are identified by matching the power behavior of the hadronic amplitude at the AdS boundary at small $z$ to the twist $\tau$ of its interpolating operator at short distances
as required by the AdS/CFT dictionary. The twist also equals the dimension of fields appearing in chiral super-multiplets;~\cite{Craig:2009rk} thus the twist of a hadron equals the number of constituents plus the relative orbital angular momentum.
One then can apply light-front holography to relate the amplitude eigensolutions  in the fifth-dimension coordinate $z$  to the LF wavefunctions in the physical space-time variable  $\zeta$.

Equation (\ref{eq:QCDLFWE}) was derived by taking the LF bound-state Hamiltonian equation of motion as the starting
point.~\cite{deTeramond:2008ht} The term $L^2/ \zeta^2$  in the  LF equation of motion  (\ref{eq:QCDLFWE})
is derived from  the reduction of the LF kinetic energy when one transforms
to two-dimensional cylindrical coordinates $(\zeta, \varphi)$,
in analogy to the $\ell(\ell+1)/ r^2$ Casimir term in Schr\"odinger theory.  One thus establishes the interpretation of $L$ in the AdS equations of motion.
The interaction terms build confinement  corresponding to
the dilaton modification of AdS space.~\cite{deTeramond:2008ht}
The duality between these two methods provides a direct
connection between the description of hadronic modes in AdS space and
the Hamiltonian formulation of QCD in physical space-time quantized
on the light-front  at fixed LF time $\tau.$

Higher spin modes follow from shifting dimensions in the AdS wave equations.
In the soft-wall
model the usual Regge behavior is found $\mathcal{M}^2 \sim n +
L$, predicting the same multiplicity of states for mesons
and baryons as observed experimentally.~\cite{Klempt:2007cp}
It is possible to extend the model to hadrons with heavy quark constituents
by introducing nonzero quark masses and short-range Coulomb
corrections.  For other
recent calculations of the hadronic spectrum based on AdS/QCD, see Refs.
~\cite{Gutsche:2011vb, deTeramond:2005su, BoschiFilho:2005yh, Evans:2006ea, Hong:2006ta, Colangelo:2007pt,  Forkel:2007cm, Forkel:2007ru, Vega:2008af, Nawa:2008xr,  dePaula:2008fp, Colangelo:2008us, Forkel:2008un, Ahn:2009px, Sui:2009xe, Kapusta:2010mf, Zhang:2010bn, Iatrakis:2010zf,  Branz:2010ub, Kirchbach:2010dm, Sui:2010ay,  Wang:2011zzc, Mady:2011ju}.

For baryons, the light-front wave equation is a linear equation
determined by the LF transformation properties of spin 1/2 states. A linear confining potential
$U(\zeta) \sim \kappa^2 \zeta$ in the LF Dirac
equation leads to linear Regge trajectories.~\cite{Brodsky:2008pg}  For fermionic modes the light-front matrix
Hamiltonian eigenvalue equation $D_{LF} \vert \psi \rangle = \mathcal{M} \vert \psi \rangle$, $H_{LF} = D_{LF}^2$,
in a $2 \times 2$ spinor  component
representation is equivalent to the system of coupled linear equations
\begin{eqnarray} \label{eq:LFDirac} \nonumber
- \frac{d}{d\zeta} \psi_- -\frac{\nu+{1\over 2}}{\zeta}\psi_-
- \kappa^2 \zeta \psi_-&=&
\mathcal{M} \psi_+, \\ \label{eq:cD2k}
  \frac{d}{d\zeta} \psi_+ -\frac{\nu+{1\over 2}}{\zeta}\psi_+
- \kappa^2 \zeta \psi_+ &=&
\mathcal{M} \psi_-,
\end{eqnarray}
with eigenfunctions
\begin{eqnarray} \nonumber
\psi_+(\zeta) &\sim& \zeta^{\frac{1}{2} + \nu} e^{-\kappa^2 \zeta^2/2}
  L_n^\nu(\kappa^2 \zeta^2) ,\\
\psi_-(\zeta) &\sim&  \zeta^{\frac{3}{2} + \nu} e^{-\kappa^2 \zeta^2/2}
 L_n^{\nu+1}(\kappa^2 \zeta^2),
\end{eqnarray}
and  eigenvalues $\mathcal{M}^2 = 4 \kappa^2 (n + \nu + 1)$.

\begin{figure}[h]
\begin{center}
\includegraphics[width=13.6cm]{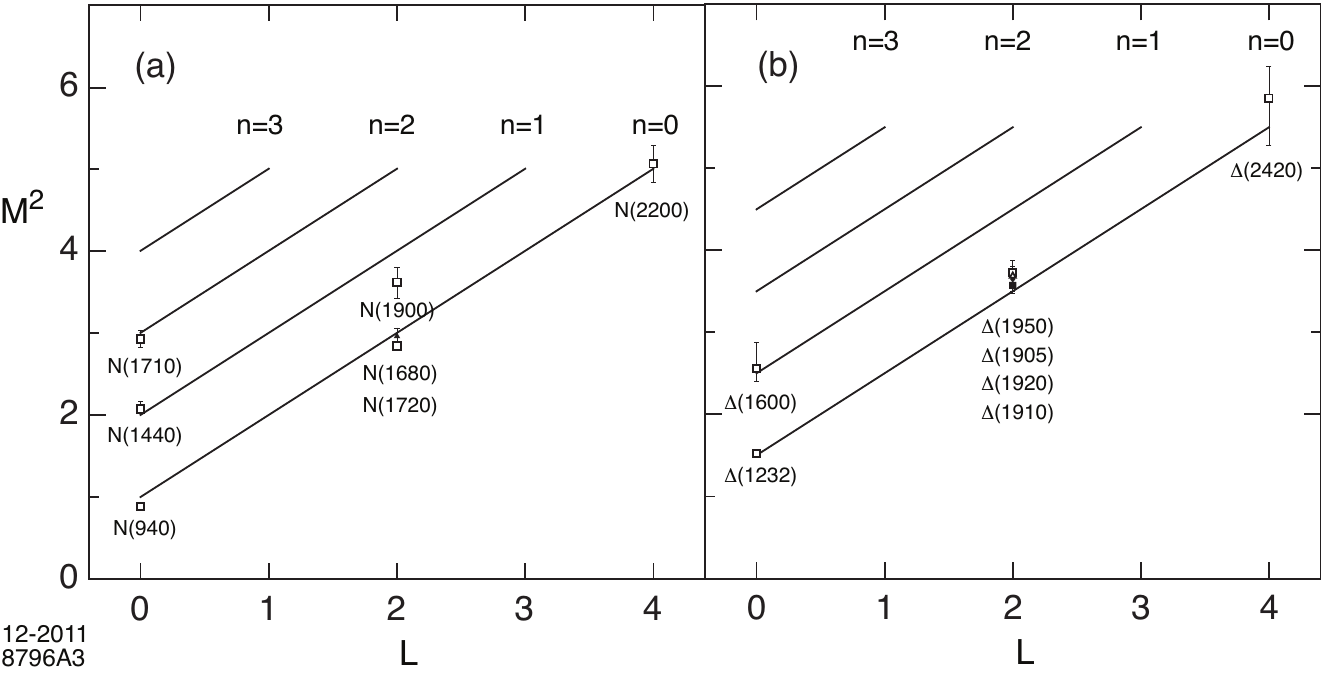}
\caption{Positive-parity Regge trajectories for  the  $N$ and $\Delta$ baryon families for $\kappa= 0.5$ GeV.}
\label{Baryons}
\end{center}
\end{figure}

The baryon interpolating operator
$ \mathcal{O}_{3 + L} =  \psi D_{\{\ell_1} \dots
 D_{\ell_q } \psi D_{\ell_{q+1}} \dots
 D_{\ell_m\}} \psi$,  $L = \sum_{i=1}^m \ell_i$, is a twist 3,  dimension $9/2 + L$ with scaling behavior given by its
 twist-dimension $3 + L$. We thus require $\nu = L+1$ to match the short distance scaling behavior. Higher spin modes are obtained by shifting dimensions for the fields.
 Thus, as in the meson sector,  the increase  in the
mass squared for higher baryonic state is
$\Delta n = 4 \kappa^2$, $\Delta L = 4 \kappa^2$ and $\Delta S = 2 \kappa^2,$
relative to the lowest ground state,  the proton. Since our starting point to find the bound state equation of motion for baryons is the light-front, we fix the overall energy scale identical for mesons and baryons by imposing chiral symmetry to the pion~\cite{deTeramond:2010we} in the LF Hamiltonian equations. By contrast, if we start with a five-dimensional action for a scalar field in presence of a positive sign dilaton, the pion is automatically massless.

The resulting predictions for the spectroscopy of positive-parity light baryons are shown in Fig. \ref{Baryons}.
As for the predictions for mesons in Fig. \ref{pionVM}, only confirmed PDG~\cite{FERMILAB-PUB-10-665-PPD} states are shown.
The Roper state $N(1440)$ and the $N(1710)$ are well accounted for in this model as the first  and second radial
states.   Likewise the $\Delta(1660)$ corresponds to the first radial state of the $\Delta$ family.  The $N(1900)$ is explained in this framework as a doubly excited  $n=1$, $L=2$ state.

The model is  successful in explaining the important parity degeneracy observed in the light baryon spectrum, such as the $L\! =\!2$, $N(1680)\!-\!N(1720)$ degenerate pair and the $L=2$, $\Delta(1905)$,  $\Delta(1910)$,  $\Delta(1920)$,  $\Delta(1950)$ states which are degenerate
within error bars.
Our results for the positive parity $\Delta$ states agree with those of Ref.~\cite{Forkel:2007cm}, but further insight is required to understand better the complex structure of the of the negative parity baryons.
Parity degeneracy of baryons is also a property of the hard wall model, but radial states are not well described in this model~\cite{deTeramond:2005su}.

\section{Hadronic Form Factors in Light-Front AdS/QCD \label{FF}}

In the higher dimensional gravity theory, the hadronic transition matrix element  corresponds to
the  coupling of an external electromagnetic field $A^M(x,z)$  for a photon propagating in AdS space with the extended field $\Phi_P(x,z)$ describing a meson in AdS is~\cite{Polchinski:2002jw}
 \begin{equation} \label{MFF}
 \int d^4x \, dz  \sqrt{g} \, A^M(x,z)
 \Phi^*_{P'}(x,z) \overleftrightarrow\partial_M \Phi_P(x,z)
  \sim
 (2 \pi)^4 \delta^4 \left( P'  \! - P - q\right) \epsilon_\mu  (P + P')^\mu F_M(q^2) .
 \end{equation}
To simplify the discussion we will first describe a  model with a wall at  $z \sim 1/\Lambda_{\rm QCD}$ -- the hard wall model -- which limits the propagation of the string modes in AdS space beyond  the IR separation $z \sim 1/\Lambda_{\rm QCD}$ and also
sets the gap scale.~\cite{Polchinski:2001tt} 
 The coordinates of AdS$_5$ are the Minkowski coordinates $x^\mu$ and $z$ labeled $x^M = (x^\mu, z)$,
 with $M, N = 1, \cdots 5$,  and $g$ is the determinant of the metric tensor. 
The pion has initial and final four momentum $P$ and $P'$ respectively and $q$ is the four-momentum transferred to the pion by the photon with polarization $\epsilon_\mu$.
The expression on the right-hand side
of (\ref{MFF}) represents the space-like QCD electromagnetic transition amplitude in physical space-time
$\langle P' \vert J^\mu(0) \vert P \rangle = \left(P + P' \right)^\mu F_M(q^2)$.
It is the EM matrix element of the quark current  $J^\mu = e_q \bar q \gamma^\mu q$, and represents a local coupling to pointlike constituents. Although the expressions for the transition amplitudes look very different, one can show  that a precise mapping of the matrix elements  can be carried out at fixed light-front time.~\cite{Brodsky:2006uqa, Brodsky:2007hb}

The form factor is computed in the light front from the matrix elements of the plus-component of the current $J^+$, in order to avoid coupling to Fock states with different numbers of constituents.
For definiteness we shall consider 
the $\pi^+$  valence Fock state 
$\vert u \bar d\rangle$
\begin{equation}  \label{eq:PiFFb}
 F_{\pi^+}(q^2)  =  2 \pi \int_0^1 \! \frac{dx}{x(1-x)}  \int \zeta d \zeta \,
J_0 \! \left(\! \zeta q \sqrt{\frac{1-x}{x}}\right) 
\left\vert \psi_{u \bar d/ \pi}\!(x,\zeta)\right\vert^2,
\end{equation}
where $\zeta^2 =  x(1  -  x) \mathbf{b}_\perp^2$ and $F_{\pi^+}(q\!=\!0)=1$. 
We now compare this result with the electromagnetic  form factor 
in  AdS  space time. The incoming electromagnetic field propagates in AdS according to
$A_\mu(x^\mu ,z) = \epsilon_\mu(q) e^{-i q \cdot x} V(q^2, z)$ in the gauge $A_z = 0$ (no physical polarizations along the AdS variable $z$).  The bulk-to-boundary propagator
 $V(q^2,z)$ is the solution of the AdS wave equation
given by ($Q^2 = - q^2 > 0$) $V(Q^2, z) = z Q K_1(z Q)$,
with boundary conditions~\cite{Polchinski:2002jw} 
$V(Q^2 \! = 0, z ) = V(Q^2, z = 0) = 1$.

The propagation of the pion in AdS space is described by a normalizable mode
$\Phi_P(x^\mu, z) = e^{-i P  \cdot x} \Phi(z)$ with invariant  mass $P_\mu P^\mu = \mathcal{M}^2$ and plane waves along Minkowski coordinates $x^\mu$.  
Extracting the overall factor  $(2 \pi)^4 \delta^4 \left( P'  \! - P - q\right)$ from momentum conservation at the vertex from integration over Minkowski variables in (\ref{MFF}) we find
\cite{Polchinski:2002jw} 
\begin{equation}
F(Q^2) = R^3 \int \frac{dz}{z^3} \, V(Q^2, z)  \, \Phi^2(z),
\label{eq:FFAdS}
\end{equation}
where $F(Q^2\! = 0) = 1$. 
We use  the integral representation of $V(Q^2,z)= \int_0^1 \! dx \, J_0 \! \left(\! z  Q \sqrt{\frac{1-x}{x}}\right)$
and the factorization of the LFWF
\begin{equation} \label{eq:psiphi}
\psi(x,\zeta, \varphi) =  X(x) \frac{\phi(\zeta)}{\sqrt{2 \pi \zeta}} ,
\end{equation}
to compare with  the light-front QCD  form factor  (\ref{eq:PiFFb}) for arbitrary values of $Q$. We find the result
$\phi(\zeta) = (\zeta/R)^{3/2} \Phi(\zeta)$ and $X(x) = \sqrt{x(1-x)}$,~\cite{Brodsky:2006uqa}
where we identify the transverse impact LF variable $\zeta$ with the holographic variable $z$,
$z \to \zeta = \sqrt{x(1-x)} \vert \ \vec b_\perp \vert$.
The factorization given in (\ref{eq:psiphi}) is a natural factorization in the light front since the
corresponding canonical generators, the longitudinal and transverse generators $P^+$ and $\vec{P}_\perp$ are kinematical generators which commute with the LF Hamiltonian generator $P^-$.

\section{Elastic Form Factor with a  Dressed Current}

The results for the elastic form factor described above correspond to a ``free''  current propagating on AdS space. It is dual to the electromagnetic
point-like current in the Drell-Yan-West
light-front formula~\cite{Drell:1969km, West:1970av} for the pion form factor.  
The DYW formula is an exact expression for the form factor. It is written as an infinite sum of an overlap of LF Fock components with an arbitrary number of constituents.
This allows one to map state-by-state to the effective gravity theory in AdS space. 
However, this mapping has the shortcoming that the multiple pole structure of the time-like form factor cannot be obtained in the time-like region unless an infinite number of Fock states is included. 
Furthermore, the moments of the form factor at  $Q^2 = 0$ diverge term-by-term; for example one obtains an infinite charge radius.~\cite{deTeramond:2011yi}  

\begin{figure}[h]
\centering
\includegraphics[angle=0,width=6.4cm]{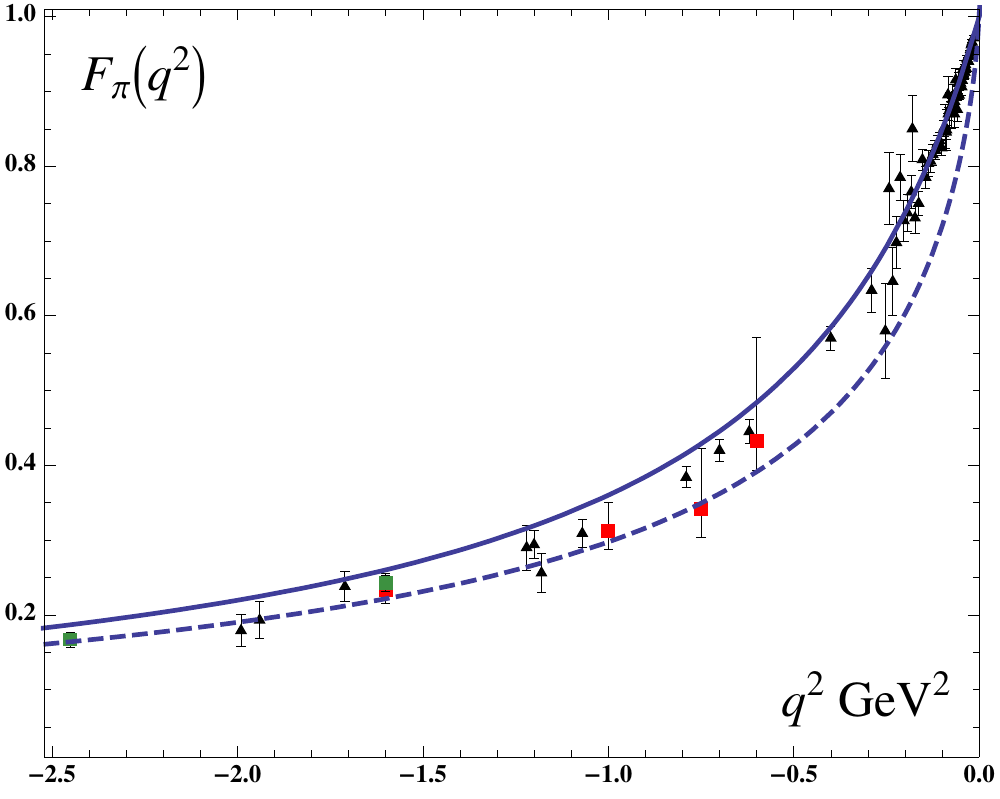} 
\caption{Space-like electromagnetic pion form factor $F_\pi(q^2)$. Continuous line: confined current, dashed  line free current. Triangles are the data compilation  from Baldini,~\cite{Baldini:1998qn} boxes  are JLAB data.~\cite{Tadevosyan:2007yd} }
\label{PionFF}
\end{figure} 

Alternatively, one can use a truncated basis of states in the LF Fock expansion with a limited number of constituents and the non-perturbative pole structure can be generated with
a  dressed EM current as in the Heisenberg picture, {\it i.e.},  the EM current becomes modified as it propagates in 
an IR deformed AdS space to simulate confinement.
The dressed current is dual to a hadronic EM current which includes any number of virtual $q \bar q$ components.  
The confined EM  current  also leads to finite moments at $Q^2=0$, as illustrated on Fig. \ref{PionFF}. 
 The pion form factor and the vector meson poles residing in the dressed current in the soft wall model require choosing  a value of $\kappa$ smaller by a factor of $1/\sqrt 2$  than the canonical value of  $\kappa$ which determines the mass scale of the hadronic spectra.  This shift is apparently due to the fact that the transverse current in $e^+ e^- \to q \bar q$ creates a quark pair with $L^z= \pm 1$ instead of the $L^z=0$ $q \bar q$ composition of the vector mesons in the spectrum.
Other recent computations of the pion form factor are given in
Refs.~\cite{Kwee:2007dd,Grigoryan:2007wn, Bayona:2010bg}.

The effective potential corresponding to a dilaton profile $\exp(\pm \kappa^2 z^2)$ has the form of a harmonic oscillator confining potential $\kappa^4 z^2$. The normalizable solution for a meson of  twist $\tau$ (the number of constituents for a given Fock component) corresponding to
 the lowest radial $n = 0$ and orbital $L=0$ state is given by
\begin{equation}  \label{eq:Phitau}
\Phi^\tau(z) =   \sqrt{\frac{2 P_{\tau}}{\Gamma(\tau \! - \! 1)} } \, \kappa^{\tau -1} z ^{\tau} e^{- \kappa^2 z^2/2},
\end{equation} 
with normalization
\begin{equation} \label{eq:PhitauNorm}
\langle\Phi^\tau\vert\Phi^\tau\rangle = \int \frac{dz}{z^3} \, e^{- \kappa^2 z^2} \Phi^\tau(z)^2  = P_\tau,
\end{equation}
where $P_\tau$ is the probability for the twist $\tau$ mode (\ref{eq:Phitau}). 
This agrees with the fact that the field $\Phi^\tau$ couples to a local hadronic interpolating operator of twist $\tau$ 
defined at the asymptotic boundary of AdS space, and thus
the scaling dimension of $\Phi^\tau$ is $\tau$. 
In the case of soft-wall potential
the EM bulk-to-boundary propagator is~\cite{Brodsky:2007hb, Grigoryan:2007my}
\begin{equation} \label{eq:Vkappa}
V(Q^2,z) = \Gamma\left(1 + \frac{Q^2}{4 \kappa^2}\right) U\left(\frac{Q^2}{4 \kappa^2}, 0, \kappa^2 z^2\right),
\end{equation}
where $U(a,b,c)$ is the Tricomi confluent hypergeometric function.
The modified current $V(Q^2,z)$ has the same boundary conditions as the free current,
and reduces to a free current in the  limit $Q^2 \to \infty$.
Eq.~(\ref{eq:Vkappa}) can be conveniently written in terms of the integral representation~\cite{Grigoryan:2007my}
\begin{equation}  \label{Vx}
V(Q^2,z) = \kappa^2 z^2 \int_0^1 \! \frac{dx}{(1-x)^2} \, x^{\frac{Q^2}{4 \kappa^2}} 
e^{-\kappa^2 z^2 x/(1-x)}.
\end{equation} 

Substituting in (\ref{eq:FFAdS}) the expression for the hadronic state (\ref{eq:Phitau}) with twist $\tau$  and the bulk-to-boundary propagator (\ref{Vx}) we find the 
elastic form factor for a twist $\tau$ Fock component  $F_\tau(Q^2)$
($Q^2 = - q^2 > 0$)~\cite{Brodsky:2007hb} 
\begin{equation} \label{Ftau}   
 F_\tau(Q^2) =  \frac{P_\tau}{{\Big(1 + \frac{Q^2}{M^2_\rho} \Big) }
 \Big(1 + \frac{Q^2}{M^2_{\rho'}}  \Big)  \cdots 
       \Big(1  + \frac{Q^2}{M^2_{\rho^{\tau-2}}} \Big)} ,
\end{equation}
which is  expressed as a $\tau - 1$ product of poles along the vector meson Regge radial trajectory.
For a pion, for example, the lowest Fock state -- the valence state -- is a twist-2 state, and thus the form factor is the well known monopole form.~\cite{Brodsky:2007hb}
The remarkable analytical form of (\ref{Ftau}),
expressed in terms of the $\rho$ vector meson mass and its radial excitations, incorporates the correct scaling behavior from the constituent's hard scattering with the photon and the mass gap from confinement. 

\begin{figure}[h]
\begin{center}
 \includegraphics[width=7.0cm]{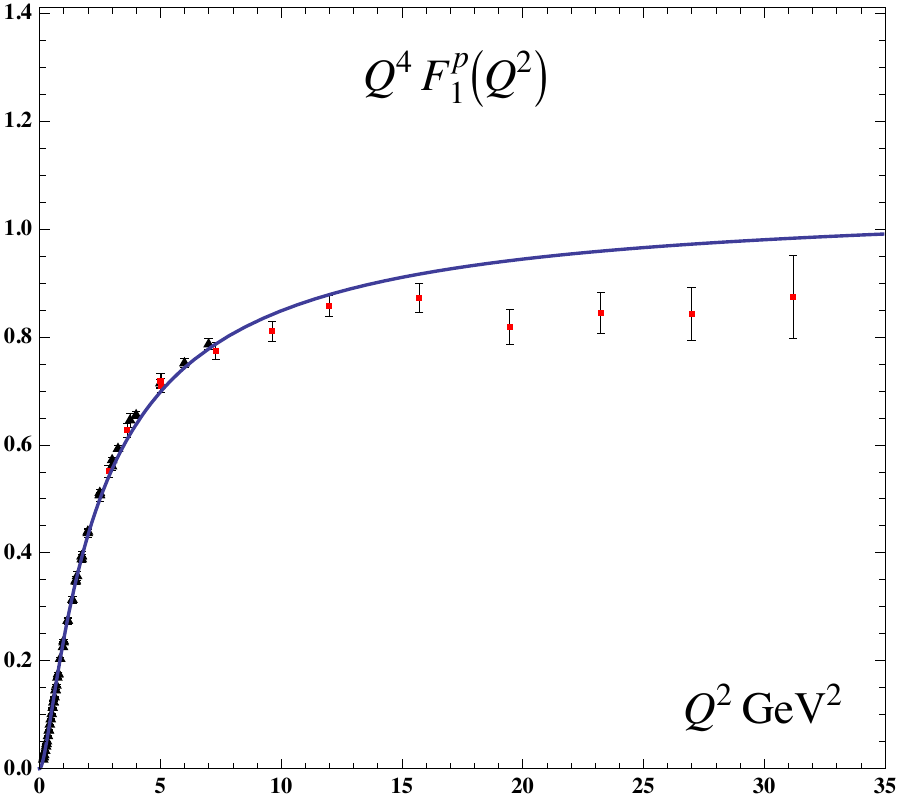}   \hspace{20pt}
\includegraphics[width=7.1cm]{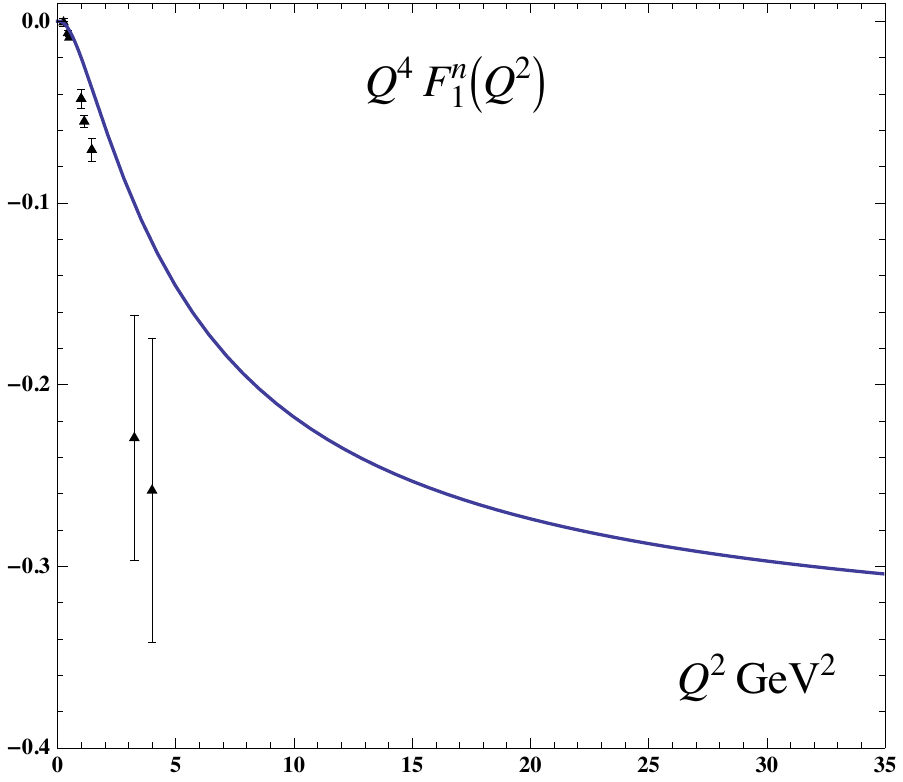}
 \caption{Predictions for $Q^4 F_1^p(Q^2)$ and $Q^4 F_1^n(Q^2)$ in the
soft wall model. Data compilation  from Diehl.~\cite{Diehl:2005wq} }
\label{fig:nucleonFF}
\end{center}
\end{figure}

As an example  of the scaling behavior of a twist $\tau = 3$ hadron, we compute the spin non-flip
nucleon form factor in the soft wall model.~\cite{Brodsky:2008pg} The proton and neutron Dirac
form factors are given by
\begin{equation}
F_1^p(Q^2) =  \! \int  d \zeta \, V(Q^2, \zeta) \,
  \vert \psi_+(\zeta)\vert^2 ,
\end{equation}
\begin{equation}
F_1^n(Q^2) =  - \frac{1}{3}  \! \int  d \zeta  \,  V(Q^2, \zeta)
 \left[\vert \psi_+(\zeta)\vert^2 - \vert\psi_-(\zeta)\vert^2\right],
 \end{equation}
where $F_1^p(0) = 1$,~ $F_1^n(0) = 0$.
Plus and minus components of the twist 3 nucleon LFWF are
\begin{equation} \label{eq:PhipiSW}
\psi_+(\zeta) \!=\! \sqrt{2} \kappa^2 \, \zeta^{3/2}  e^{-\kappa^2 \zeta^2/2},  ~~~
\Psi_-(\zeta) \!=\!  \kappa^3 \, \zeta^{5/2}  e^{-\kappa^2 \zeta^2/2}.
\end{equation}
The results for $Q^4 F_1^p(Q^2)$ and $Q^4 F_1^n(Q^2)$  are shown in
Fig. \ref{fig:nucleonFF}. To compare with physical data we have shifted the poles in expression (\ref{Ftau}) to their physical values located at $M^2 = 4 \kappa^2(n + 1/2)$.

\subsection{Effective Light-Front Wave Function From Holographic Mapping of a Confined Electromagnetic Current }

It is also possible to find a precise mapping of a confined EM current propagating in a warped AdS space to the light-front QCD Drell-Yan-West expression for the form factor. In this case we we find an effective LFWF, which corresponds to a superposition of an infinite number of Fock states generated by the ``dressed'' confined current. For the soft-wall model this mapping can be done analytically.   

The form factor in light-front  QCD can be expressed in terms of an effective single-particle density~\cite{Soper:1976jc}
\begin{equation} 
F(Q^2) =  \int_0^1 dx \, \rho(x,Q),
\end{equation}
where
\begin{equation} \label{rhoQCD}
\rho(x, Q) = 2 \pi \int_0^\infty \!  b \,  db \, J_0(b Q (1-x)) \vert \psi(x,b)\vert^2,
\end{equation}
for a two-parton state ($b = \vert {b}_\perp \vert$).

 We can also compute an effective density on the gravity side corresponding to a twist $\tau$ hadronic mode $\Phi_\tau$ in a modified AdS space.
 For the soft-wall model the expression is~\cite{Brodsky:2007hb}
 \begin{equation}  \label{rhoAdS}
\rho(x,Q) = (\tau \!-\!1) \, (1 - x)^{\tau-2} \, x^{\frac{Q^2}{4 \kappa^2}} .
\end{equation}
To compare (\ref{rhoAdS}) with the QCD expression (\ref{rhoQCD})  we use the integral
$\int_0^\infty \! u \, du  \, J_0(\alpha u) \,e^{- \beta u^2} = \frac{1}{2 \beta} \, e^{-\alpha^2/4\beta},$
and the relation $x^\gamma  = e^{\gamma \ln(x)}$. We find the effective two-parton  LFWF  ($\tau = 2$)
\begin{equation} \label{ELFWF} 
\psi(x, {b}_\perp) = \kappa \frac{ (1-x)}{\sqrt{\pi \ln(\frac{1}{x})}} \,
e^{- {1\over 2} \kappa^2 {b}_\perp^2  (1-x)^2 / \ln(\frac{1}{x})}
\end{equation}
in impact space. The momentum space expression follows from the Fourier transform of  (\ref{ELFWF})
and it is given by  
\begin{equation} 
\psi(x, {k}_\perp) 
 = 4 \pi \, \frac{ \sqrt{\ln\left(\frac{1}{x}\right)}}{\kappa (1-x)} \,
e^{ - {k}_\perp^2/2 \kappa^2 (1-x)^2 \ln \left(\frac{1}{x}\right)}.
\end{equation}
The effective LFWF  encodes  non-perturbative dynamical aspects that cannot be learned from a term-by-term holographic mapping, unless one adds an infinite number of terms.  Furthermore, it has the right analytical properties to reproduce the bound state vector meson pole in the time-like EM form factor. Unlike the ``true'' valence LFWF, the effective LFWF, which represents a sum of an infinite number of Fock components, is not symmetric in the longitudinal variables $x$ and $1-x$ for the active and spectator quarks,  respectively.  
 Holographic QCD methods have also been used to obtain generalized parton distributions (GPDs) in
 Refs.~\cite{Vega:2010ns} and  \cite{Nishio:2011xa}. Nucleon transition form factors have been studied in  
 Refs.~\cite{deTeramond:2011qp}    and   \cite{Grigoryan:2009pp},   and structure functions is Ref.~\cite{Bayona:2011xj}.

\section{Higher Fock Components in Light-Front Holographic QCD}

The five-quark Fock state of the proton's LFWF $ \vert uud Q \bar Q \rangle$ is the primary origin of the sea quark distributions of the proton.~\cite{Brodsky:1996hc,Brodsky:2000sk}  Experiments show that the sea quarks have remarkable nonperturbative features, such as $\bar u(x) \ne \bar d(x)$, an intrinsic strangeness~\cite{Airapetian:2008qf} distribution $s(x)$ appearing at $x > 0.1$, as well as intrinsic charm and bottom distributions at large $x$.  Such distributions~\cite{Brodsky:1984nx,Franz:2000ee} will arise rigorously from $g g \to Q \bar Q \to gg $ insertions connected to the valence quarks in the proton self-energy; in fact, they fit a universal intrinsic quark model,~\cite{Brodsky:1980pb} as shown by Chang and Peng.~\cite{Chang:2011du}

The LF Hamiltonian eigenvalue equation (\ref{LFH}) is a matrix in Fock space which represents an infinite number of coupled integral equations for the Fock components $\psi_n = \langle n \vert \psi \rangle$. The resulting potential in quantum field theory can be considered as an instantaneous four-point effective interaction 
in LF time, similar to the instantaneous gluon exchange in the light-cone gauge $A^+ = 0$,
which leads
to $q q \to qq$, $q \bar q \to q \bar q$, $ q \to q q \bar q$ and $\bar q \to \bar q q \bar q$ as in QCD(1+1). 
Thus in holographic QCD, higher Fock states can have any number of extra $q \bar q$ pairs, but surprisingly no dynamical gluons.
This unusual property of AdS/QCD may explain the dominance of quark interchange~\cite{Gunion:1972qi}
 over quark annihilation or gluon exchange contributions in large angle elastic scattering.~\cite{264273} This result
 is consistent with the flux-tube interpretation of QCD~\cite{Print-84-0830}  where soft gluons interact so strongly that they are sublimated into a color confinement potential for quarks. 
 In Sec. \ref{SG} we discuss further experimental results in hadron physics which support this picture.
 
  \begin{figure}[h]
\centering
\includegraphics[width=6.45cm]{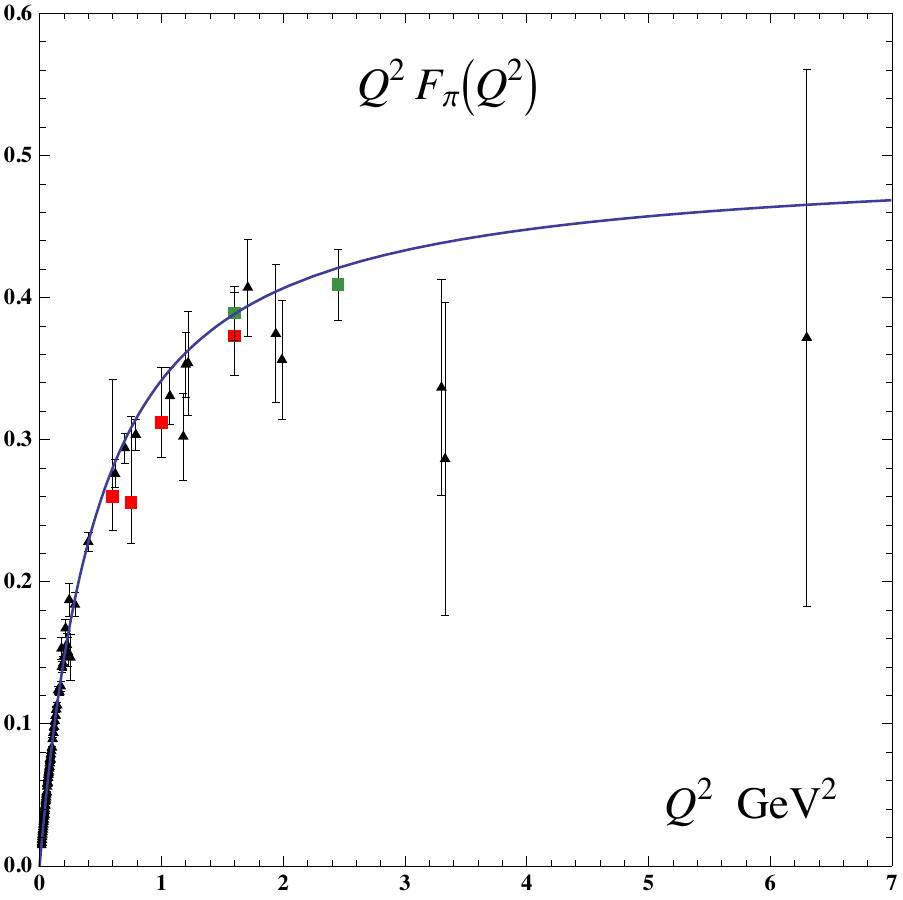} \hspace{8pt}
\includegraphics[width=7.10cm]{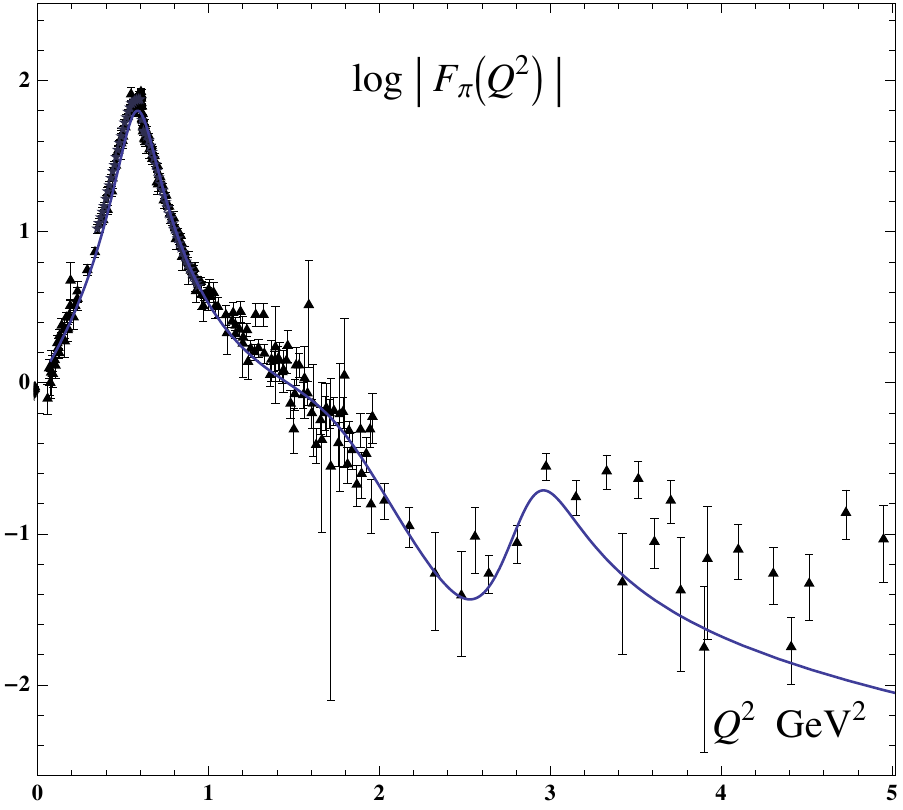}
\caption{Structure of the space- and time-like pion form factor in light-front holography for a truncation of the pion wave function up to twist-four.
Triangles are the data compilation  from Baldini  {\it et al.},~\cite{Baldini:1998qn}  squares are JLAB data.~\cite{Tadevosyan:2007yd} }
\label{pionFFhfs}
\end{figure}

To illustrate the relevance of higher Fock states and the absence of dynamical gluons at the hadronic scale, we discuss  a simple semi-phenomenological model of the elastic form factor of the pion where we include the first two components in a Fock expansion of the pion wave function
$\vert \pi \rangle  = \psi_{q \bar q /\pi} \vert q \bar q  \rangle_{\tau=2}
+  \psi_{q \bar q q \bar q} \vert q \bar q  q \bar q  \rangle_{\tau=4} + \cdots$ ,
where the $J^{PC} = 0^{- +}$ twist-two and twist-4 states $\vert q \bar q \rangle$  and  $\vert q \bar q q \bar q  \rangle$ are created by the interpolating operators
$\bar q \gamma^+ \gamma_5  q$ and $ \bar q \gamma^+ \gamma_5  q  \bar q q$ respectively. 

Since the charge form factor is a diagonal operator, the final expression for the form factor corresponding to the truncation up to twist four is the sum of two terms, a monopole and a three-pole term.
In the strongly coupled semiclassical gauge/gravity limit, hadrons have zero widths and are stable. One can nonetheless modify the formula (\ref{Ftau}) to introduce a finite width:
$q^2 \to q^2 + \sqrt 2 i M \Gamma$.  We choose the values $\Gamma_\rho =  140$ MeV,   $\Gamma_{\rho'} =  360$ MeV and  $\Gamma_{\rho''} =  120$ MeV.  The results for the pion form factor with twist two and four Fock states  are shown in Fig. \ref{pionFFhfs}. The results correspond to $P_{q \bar q q \bar q}$ = 13 \%, the admixture of the
$\vert q \bar q q \bar q  \rangle$ state. The value of $P_{q \bar q q \bar q}$ (and the widths) are input in the model. The value of $\kappa$ is determined from the $\rho$ mass and the masses of the radial excitations follow from 
$M^2 \to 4 \kappa^2(n + 1/2)$. The time-like structure of the pion form factor displays a rich pole structure with constructive and destructive interferences, which is incompatible with the admixture of the twist three state $\vert q \bar q g \rangle$ containing a dynamical gluon as the interference becomes opposite in sign.

\section{Nonperturbative Running Coupling from Light-Front Holography \label{alphaAdS}}

The concept of a running coupling $\alpha_s(Q^2)$  in QCD is usually restricted to the perturbative domain.  However, as in QED, it is useful to define the coupling as an analytic function valid over the full space-like and time-like domains.
The study of the non-Abelian QCD coupling at small momentum transfer is a complex problem because of  gluonic self-coupling and color confinement.

The definition of the running coupling in perturbative quantum field theory is scheme-dependent.  As discussed by Grunberg,~\cite{Grunberg} an effective coupling or charge can be defined directly from physical observables.
Effective charges defined from
different observables can be related  to each other in the leading-twist domain using commensurate scale relations
(CSR).~\cite{CSR}   The  potential between infinitely heavy quarks can be defined analytically in momentum transfer
 space as the product  of the running coupling times the Born gluon propagator: $V(q)  = - 4 \pi C_F {\alpha_V(q) / q^2}$.   This effective charge defines a renormalization scheme -- the $\alpha_V$ scheme of Appelquist, Dine,  and Muzinich.~\cite{Appelquist:1977tw}
In fact, the holographic coupling $\alpha_s^{AdS}(Q^2)$ can be considered to be the nonperturbative extension of the
$\alpha_V$ effective charge defined in Ref. \cite{Appelquist:1977tw}.
We can also make  use of the $g_1$ scheme, where the strong coupling $\alpha_{g_1}(Q^2)$ is determined from
the Bjorken sum rule.~\cite{BjorkenSR} The coupling $\alpha_{g_1}(Q^2)$ has the advantage that it is the best-measured effective charge, and it can be used to extrapolate the definition of the effective coupling to large distances.~\cite{Deur:2009hu} Since $\alpha_{g_1}$ has been measured at intermediate energies, it is
particularly useful for studying  the transition from
small to large distances.

We have  shown with Deur~\cite{Brodsky:2010ur}   how the LF holographic mapping of effective classical gravity in AdS space, modified by a positive-sign dilaton background, can  be used to identify an analytically simple  color-confining
non-perturbative effective coupling $\alpha_s^{AdS}(Q^2)$ as a function of the space-like momentum transfer $Q^2 = - q^2$.   This coupling incorporates  confinement
and agrees well with effective charge observables and lattice simulations.
It also exhibits an infrared fixed point at small $Q^2$ and asymptotic freedom at large $Q^2$. However, the fall-off   of
$\alpha_s^{AdS}(Q^2)$  at large $Q^2$ is exponential: $\alpha_s^{AdS}(Q^2) \sim e^{-Q^2 /  \kappa^2}$, rather than the
pQCD logarithmic fall-off.    It agrees with hadron physics data
extracted phenomenologically from different observables, as well as with  the predictions of models with built-in confinement  and lattice simulations. We
also show that a phenomenological extended coupling can be defined which implements the pQCD behavior.
The  $\beta$-function derived from light-front holography becomes significantly  negative in the non-perturbative regime $Q^2 \sim \kappa^2$, where it reaches a minimum, signaling the transition region from the infrared (IR) conformal region, characterized by hadronic degrees of freedom,  to a pQCD conformal ultraviolet (UV)  regime where the relevant degrees of freedom are the quark and gluon constituents.  The  $\beta$-function vanishes at large $Q^2$ consistent with asymptotic freedom, and it vanishes at small $Q^2$ consistent with an infrared fixed point~\cite{Brodsky:2008be, Cornwall:1981zr}.

Let us consider a five-dimensional gauge field $F$ propagating in AdS$_5$ space in presence of a dilaton background
$\varphi(z)$ which introduces the energy scale $\kappa$ in the five-dimensional action.
At quadratic order in the field strength the action is
\begin{equation}
S =  - {1\over 4}\int \! d^5x \, \sqrt{g} \, e^{\varphi(z)}  {1\over g^2_5} \, F^2,
\label{eq:action}
\end{equation}
where the metric determinant of AdS$_5$ is $\sqrt g = ( {R/z})^5$,  $\varphi=  \kappa^2 z^2$ and the square of the coupling $g_5$ has dimensions of length.   We  can identify the prefactor
\begin{equation} \label{eq:flow}
g^{-2}_5(z) =  e^{\varphi(z)}  g^{-2}_5 ,
\end{equation}
in the  AdS  action (\ref{eq:action})  as the effective coupling of the theory at the length scale $z$.
The coupling $g_5(z)$ then incorporates the non-conformal dynamics of confinement. The five-dimensional coupling $g_5(z)$
is mapped,  modulo a  constant, into the Yang-Mills (YM) coupling $g_{YM}$ of the confining theory in physical space-time using light-front holography. One  identifies $z$ with the invariant impact separation variable $\zeta$ which appears in the LF Hamiltonian:
$g_5(z) \to g_{YM}(\zeta)$. Thus
\begin{equation}  \label{eq:gYM}
\alpha_s^{AdS}(\zeta) = g_{YM}^2(\zeta)/4 \pi \propto  e^{-\kappa^2 \zeta^2} .
\end{equation}

In contrast with the 3-dimensional radial coordinates of the non-relativistic Schr\"o-dinger theory, the natural light-front
variables are the two-dimensional cylindrical coordinates $(\zeta, \phi)$ and the
light-cone fraction $x$. The physical coupling measured at the scale $Q$ is the two-dimensional Fourier transform
of the  LF transverse coupling $\alpha_s^{AdS}(\zeta)$  (\ref{eq:gYM}). Integration over the azimuthal angle
 $\phi$ gives the Bessel transform
 \begin{equation} \label{eq:2dimFT}
\alpha_s^{AdS}(Q^2) \sim \int^\infty_0 \! \zeta d\zeta \,  J_0(\zeta Q) \, \alpha_s^{AdS}(\zeta),
\end{equation}
in the $q^+ = 0$ light-front frame where $Q^2 = -q^2 = - {\vec q}_\perp^2 > 0$ is the square of the space-like
four-momentum transferred to the
hadronic bound state.   Using this ansatz we then have from  Eq.  (\ref{eq:2dimFT})
\begin{equation}
\label{eq:alphaAdS}
\alpha_s^{AdS}(Q^2) = \alpha_s^{AdS}(0) \, e^{- Q^2 /4 \kappa^2}.
\end{equation}
In contrast, the negative dilaton solution $\varphi=  -\kappa^2 z^2$ leads to an integral which diverges at large $\zeta$.
We identify $\alpha_s^{AdS}(Q^2)$ with the physical QCD running coupling
in its nonperturbative domain.

The flow equation  (\ref{eq:flow}) from the scale dependent measure for the gauge fields can be understood as a consequence of field-strength renormalization.
In physical QCD we can rescale the non-Abelian gluon field  $A^\mu \to \lambda A^\mu$  and field strength
$G^{\mu \nu}  \to \lambda G^{\mu \nu}$  in the QCD Lagrangian density  $\mathcal{L}_{\rm QCD}$ by a compensating rescaling of the coupling strength $g \to \lambda^{-1} g.$  The renormalization of the coupling $g _{phys} = Z^{1/2}_3  g_0,$   where $g_0$ is the bare coupling in the Lagrangian in the UV-regulated theory,  is thus  equivalent to the renormalization of the vector potential and field strength: $A^\mu_{ren} =  Z_3^{-1/2} A^\mu_0$, $G^{\mu \nu}_{ren} =  Z_3^{-1/2} G^{\mu \nu}_0$   with a rescaled Lagrangian density
${\cal L}_{\rm QCD}^{ren}  = Z_3^{-1}  { \cal L}_{\rm QCD}^0  = (g_{phys}/g_0)^{-2}  \mathcal{L}_0$.
 In lattice gauge theory,  the lattice spacing $a$ serves as the UV regulator, and the renormalized QCD coupling is determined  from the normalization of the gluon field strength as it appears  in the gluon propagator. The inverse of the lattice size $L$ sets the mass scale of the resulting running coupling.
As is the case in lattice gauge theory, color confinement in AdS/QCD reflects nonperturbative dynamics at large distances. The QCD couplings defined from lattice gauge theory and the soft wall holographic model are thus similar in concept, and both schemes are expected to have similar properties in the nonperturbative domain, up to a rescaling of their respective momentum scales.

\begin{figure}[h]
\begin{center}
\includegraphics[width=6.8cm]{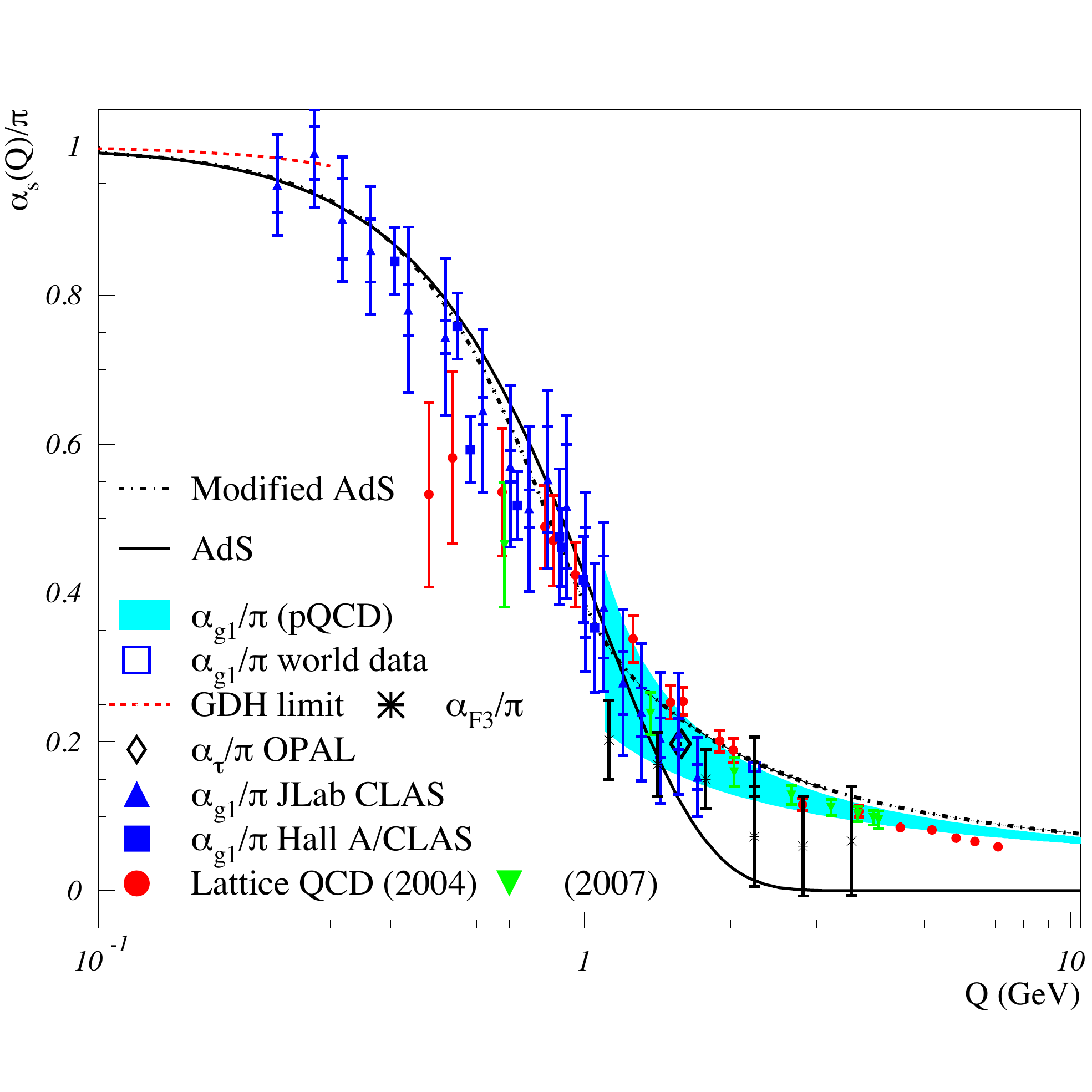} ~~~
\includegraphics[width=6.75cm]{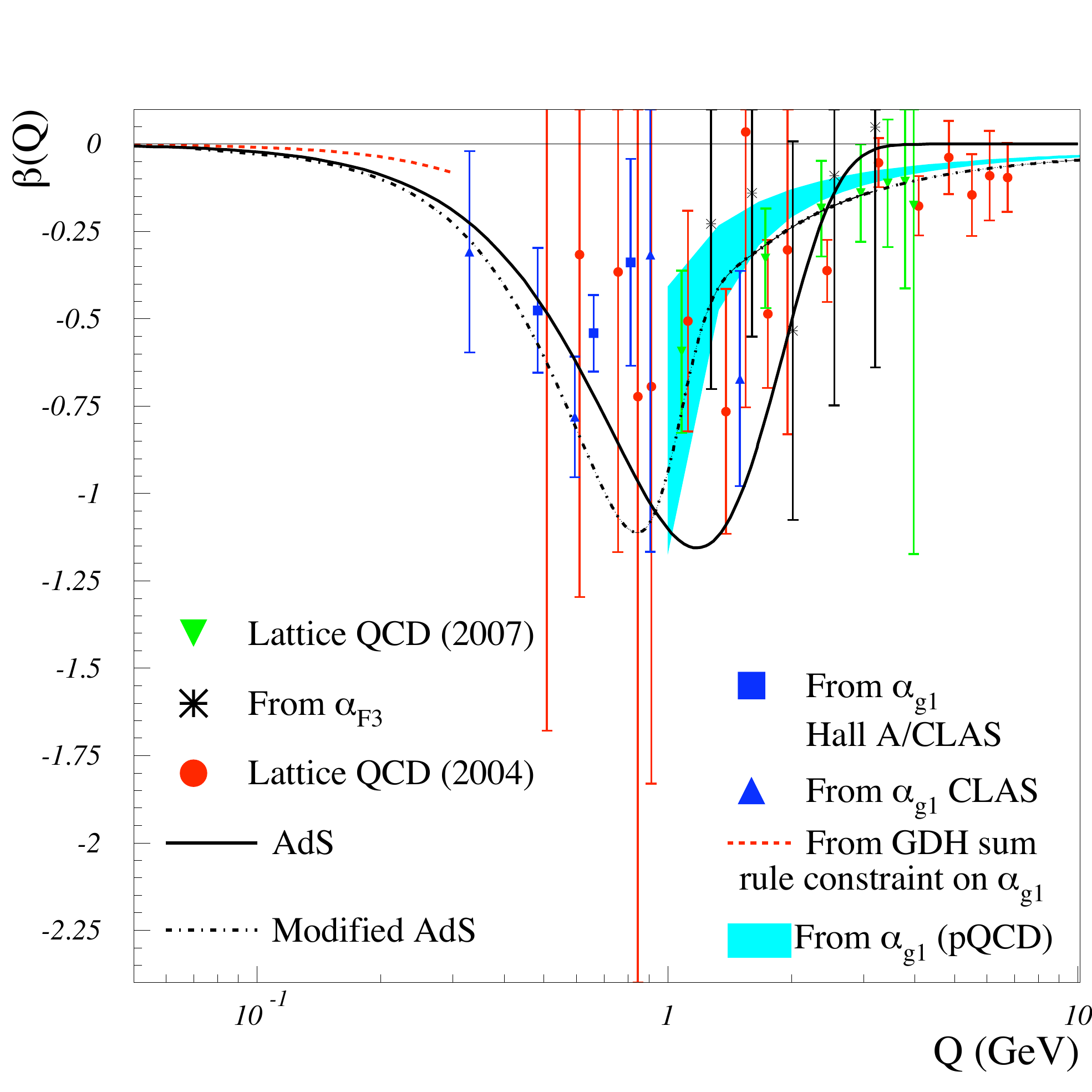}
 \caption{(a) Effective coupling  from LF holography  for  $\kappa = 0.54 ~ {\rm GeV}$ compared with effective QCD couplings  extracted from
different observables and lattice results. (b) Prediction for the $\beta$-function compared to  lattice simulations,  JLab and CCFR results  for the Bjorken sum rule effective charge.}
\label{alphas}
\end{center}
\end{figure}

The effective coupling  $\alpha^{AdS}(Q^2)$ (solid line) is compared in Fig. \ref{alphas} with  experimental and lattice data. For this comparison to be meaningful, we have to impose the same normalization on the AdS coupling as the $g_1$ coupling. This defines $\alpha_s^{AdS}$ normalized to the $g_1$ scheme: $\alpha_{g_1}^{AdS}\left(Q^2 \! =0\right) = \pi.$
Details on the comparison with other effective charges are given in Ref. ~\cite{Deur:2005cf}.

The couplings in Fig. \ref{alphas} (a) agree well in the strong coupling regime  up to $Q  \! \sim \! 1$ GeV.  The value $\kappa  = 0.54 ~ {\rm GeV}$ is determined from the vector meson Regge trajectory.~\cite {deTeramond:2009xk}
The lattice results shown in Fig. \ref{alphas} from Ref.~\cite{Furui:2006py} have been scaled to match the perturbative UV domain. The effective charge $\alpha_{ g_1}$ has been determined in Ref.~\cite{Deur:2005cf} from several experiments.  Fig. \ref{alphas} also displays other couplings from different observables as well as $\alpha_{g_1}$ which is computed from the
Bjorken sum rule~\cite{BjorkenSR}  over a large range of momentum transfer. 
At $Q^2\!=\!0$ one has the constraint  on the slope of $\alpha_{g_1}$ from the Gerasimov-Drell-Hearn (GDH) sum rule~\cite{GDH} which is also shown in the figure.
The results show no sign of a phase transition, cusp, or other non-analytical behavior, a fact which allows us to extend the functional dependence of the coupling to large distances.
As discussed below,
the smooth behavior of the  AdS strong
coupling also allows us to extrapolate its form to the perturbative domain.~\cite{Brodsky:2010ur}

The hadronic model obtained from the dilaton-modified AdS space provides a semi-classical first approximation to QCD.  Color confinement is introduced by the harmonic oscillator potential, but effects from gluon creation and absorption are not included in this effective theory.  The nonperturbative  confining effects vanish exponentially at large momentum transfer (Eq. (\ref{eq:alphaAdS})), and thus the logarithmic fall-off from pQCD quantum loops will dominate in this regime.
 Since the strong coupling is an analytical function of the momentum transfer at all scales, we can extend  the range of applicability of $\alpha_s^{AdS}$ by matching to a  perturbative coupling
at the transition scale  $Q \sim 1$ GeV, where pQCD contributions become important, as described in Ref. ~\cite{Brodsky:2010ur}.
The smoothly extrapolated result (dot-dashed line) for  $\alpha_{s}$ is also shown on Fig.~\ref{alphas}.
In order to have a fully analytical model, we write
\begin{equation}
\label{eq:alphafit}
\alpha_{Modified, g_1}^{AdS}(Q^2) = \alpha_{g_1}^{AdS}(Q^2) g_+(Q^2 ) + \alpha_{g_1}^{fit}(Q^2) g_-(Q^2),
\end{equation}
where $g_{\pm}(Q^2) = 1/(1+e^{\pm \left(Q^2 - Q^2_0\right)/\tau^2})$ are smeared step functions which match the two
regimes. The parameter $\tau$ represents the width of the
transition region.    Here $\alpha_{g_1}^{AdS}$ is given by Eq. (\ref{eq:alphaAdS}) with the normalization  $\alpha_{g_1}^{AdS}(0)=\pi$
-- the plain black line in Fig.~\ref{alphas} -- and $\alpha_{g_{1}}^{fit}$ in Eq. (\ref{eq:alphafit}) is the analytical
fit to the measured coupling $\alpha_{g_1}$~\cite{Deur:2005cf}.
The couplings  are chosen to have the same  normalization at $Q^2=0.$
The smoothly extrapolated result (dot-dashed line) for  $\alpha_{s}$ is also shown on Fig.~\ref{alphas}. We use the
parameters $Q_{0}^{2}=0.8$ GeV$^{2}$ and $\tau^2=0.3$ GeV$^{2}$.

\subsection{The $\beta$-Function from AdS/QCD}

The $\beta$-function for the nonperturbative  effective coupling obtained from the LF holographic mapping in a positive dilaton modified AdS background  is
\begin{equation} \label{eq:beta}
\beta^{AdS}(Q^2)  = {d \over d \log{Q^2}}\alpha^{AdS}(Q^2) = - {\pi Q^2\over 4 \kappa^2} e^{-Q^2/(4 \kappa^2)}.
\end{equation}
The solid line in Fig. \ref{alphas} (b) corresponds to the light-front holographic result Eq.  (\ref{eq:beta}).    Near $Q_0 \simeq 2 \kappa \simeq 1$ GeV, we can interpret the results as a transition from  the nonperturbative IR domain to the quark and gluon degrees of freedom in the perturbative UV  regime. The transition momentum  scale $Q_0$  is compatible with the momentum transfer for the onset of scaling behavior in exclusive reactions where quark counting rules are observed.~\cite{Brodsky:1973kr,Matveev:1973ra}
For example, in deuteron photo-disintegration the onset of scaling corresponds to  momentum transfer  of  1.0  GeV to the nucleon involved.~\cite{Gao:2004zh}  Dimensional counting is built into the AdS/QCD soft and hard wall models since the AdS amplitudes $\Phi(z)$ are governed by their twist scaling behavior $z^\tau$ at short distances, $ z \to 0$.~\cite{Polchinski:2001tt}

Also shown on Fig. \ref{alphas} (b) are the $\beta$-functions obtained from phenomenology and lattice calculations. For clarity, we present only the LF holographic predictions, the lattice results from,~\cite{Furui:2006py} and the
experimental data supplemented by the relevant sum rules.
The dot-dashed curve corresponds to the
extrapolated approximation obtained by matching to AdS results to the perturbative coupling~\cite{Brodsky:2010ur}
given by Eq. (\ref{eq:alphafit}).
The $\beta$-function extracted from LF holography, as well as the forms obtained from
the works of Cornwall,~\cite{Cornwall:1981zr} Bloch, Fisher {\it et al.},~\cite{S-D} Burkert and Ioffe~\cite{Burkert-Ioffe} and Furui and Nakajima,~\cite{Furui:2006py} are seen to have a similar shape
and magnitude.

Judging from these results, we infer that the   actual  $\beta$-function of QCD will extrapolate between the non-perturbative results for $Q < 1$ GeV and the pQCD results
for $Q > 1$ GeV. We also observe that the general conditions
\begin{eqnarray}
& \beta(Q \to 0) =  \beta(Q \to \infty) = 0 , \label{a} \\
&  \beta(Q)  <  0, ~ {\rm for} ~  Q > 0 , \label{b}\\
& \frac{d \beta}{d Q} \big \vert_{Q = Q_0}  = 0, \label{c} \\
& \frac{d \beta}{d Q}   < 0, ~ {\rm for} ~ Q < Q_0, ~~
 \frac{d \beta}{d Q}   > 0, ~ {\rm for} ~ Q > Q_0 \label{d} .
\end{eqnarray}
are satisfied by our model $\beta$-function obtained from LF holography.

Eq. (\ref{a}) expresses the fact that  QCD approaches a conformal theory in both the far ultraviolet and deep infrared regions. In the semiclassical approximation to QCD  without particle creation or absorption,
the $\beta$-function is zero and the approximate theory is scale  invariant
in the limit of massless quarks.~\cite{Parisi:1972zy} When quantum corrections are included,
the conformal behavior is
preserved at very large $Q$ because of asymptotic freedom and near $Q \to 0$ because the theory develops a fixed
point.  An infrared fixed point is in fact a natural consequence of color confinement~\cite{Cornwall:1981zr}
since the propagators of the colored fields have a maximum wavelength,  all loop
integrals in the computation of  the gluon self-energy decouple at $Q^2 \to 0$.~\cite{Brodsky:2008be} Condition (\ref{b}) for $Q^2$ large, expresses the basic anti-screening behavior of QCD where the strong coupling vanishes. The $\beta$-function in QCD is essentially negative, thus the coupling increases monotonically from the UV to the IR where it reaches its maximum value:  it has a finite value for a theory with a mass gap. Equation (\ref{c}) defines the transition region at $Q_0$ where the 
$\beta$-function has a minimum.  Since there is only one hadronic-partonic transition, the minimum is an absolute minimum; thus the additional conditions expressed in Eq (\ref{d}) follow immediately from
Eqs.  (\ref{a}-\ref{c}). The conditions given by Eqs.  (\ref{a}-\ref{d}) describe the essential
behavior of the full $\beta$-function for an effective QCD coupling whose scheme/definition is similar to that of the $V$-scheme.

\section{Sublimated Gluons  \label{SG}}

As we have discussed above,  an important feature of  AdS/QCD is its prediction for the running coupling constant in the infrared domain: 
$\alpha_s^{\rm AdS/QCD}(Q^2) \propto e^{-Q^2/4 \kappa^2}$.~\cite{Brodsky:2010ur}
The Gaussian fall-off of the AdS/QCD prediction for $\alpha_s^{\rm AdS/QCD} $ is clearly incorrect  in the short-distance domain $Q^2 > 4 \kappa^2 \simeq 1 ~{\rm GeV^2}$ where the asymptotic freedom property of QCD becomes evident, as seen from the fall-off of the measured $g_1$ effective charge $\alpha_s^{g_1}(Q^2)$.~\cite{Deur:2005cf}
We can understand the breakdown of the AdS/QCD prediction at hard scales as evidence for the appearance of dynamical gluon degrees of freedom in the $Q^2 >  {\rm GeV^2}$ domain. However, gluons with smaller virtuality are sublimated in terms of the effective confining potential.

An essential prediction of quantum chromodynamics is the existence of color-octet spin-1 gluon quanta.  If one quantizes QCD using light-front quantization and $A^+ \! =0$ light-cone gauge, then the gluon quanta have positive metric and physical polarization $S^z=\pm 1$.  Gluon jets are clearly seen in hard QCD processes,  such as the three-jet events in electron-positron annihilation $e^+e^- \to q \bar q g$.~\cite{DESY 79/53}  Empirically, the rapidity plateau of a gluon jet is higher than that of a quark jet by the factor $\frac{C_A}{C_F} = \frac{9}{4}$, as predicted at leading order in QCD. The existence of asymptotic freedom at large $Q^2$:
$\alpha_s(Q^2) \simeq  {4\pi /\beta_o\log{Q^2\over \Lambda^2_{QCD}}}$, the DGLAP evolution of structure functions, and the ERBL evolution of distribution amplitudes are all based on the existence of hard gluons.
However, empirical evidence confirming gluonic degrees of freedom at  small virtualities is lacking. For example,
\begin{enumerate}

\item 
Finding clear evidence for  $gg$ and $ggg$ gluonium bound states has been difficult., even in the ``gluon factory''  reaction  $J/\psi \to \gamma gg$~\cite{Klempt:2007cp, Brodsky:1977du}.  
Similarly, there is no clear evidence for $q \bar q g$ hybrid states.~\cite{Klempt:2007cp}

\item
One would normally expect to see gluon exchange dominate
large-angle elastic scattering
exclusive hadron-hadron scattering reactions.  In fact, as shown by Carroll {\it et al.},~\cite{264273} two body scattering amplitudes at fixed $\theta_{cm}$ are dominated by quark exchange and interchange amplitudes~\cite{Gunion:1972qi}  
rather than gluon exchange contributions.    As shown by Landshoff,~\cite{Landshoff:1974ew}  gluon exchange implies that large-angle proton-proton elastic scattering would be dominated by a sequence of three $qq \to qq$ amplitudes each with small gluon virtuality $t/9$.  The Landshoff mechanism predicts
${d\sigma\over dt}(p p \to pp) \propto {1\over t^8}$; in fact measurements are consistent with 
${d\sigma\over dt}(p p \to pp) \propto {1\over s^2 u^4 t^4}$, as predicted by quark-interchange.~\cite{Gunion:1972qi, Brodsky:1974vy}

\item 
There is no solid evidence for the existence of the Odderon -- the three-gluon exchange trajectory.

\item 
Experiments find that the $J/\psi \to \rho \pi$ is the largest two-body mode where $\psi^\prime$ almost never decays to $\rho \pi$.    
The infamous $J/\psi \to \rho \pi$  puzzle shows that the usual Zweig picture of $c \bar c$ annihilation to three gluons, each with virtuality $q^2 \simeq {M^2_{\psi}/ 9}$ is incorrect.  However, this  puzzling decay pattern is consistent with a mechanism~\cite{Brodsky:1997fj} where the $c \bar c$ of the quarkonium state flows into the $\vert q \bar q c \bar c \rangle$ Fock state of one of the final-state mesons.  The suppression of $\psi^\prime \to \rho \pi$ is then due to the node in the excited quarkonium radial wavefunction.

\end{enumerate}

These striking phenomenological features are consisted with the AdS/QCD prediction that gluon quanta with virtuality below $Q^2 \simeq 1\, {\rm GeV}^2$ are physically absent.  Thus the physical effects of gluons are evidently  {\it sublimated}, replaced by  an effective 
potential which confines quarks. The AdS/QCD picture is consistent with string descriptions of confinement and the Isgur-Paton flux tube model.~\cite{Print-84-0830}   We also note that  in (1+1) QCD quantized in light-cone gauge -- gluon quanta are absent; only the  confining potential $\frac{1}{{k^+}^2}$ which is instantaneous in light-front time $\tau = t +z/c$  remains.

\section{The Principle of Maximal Conformality}

A key problem in making precise perturbative QCD predictions is
the uncertainty in determining the renormalization scale $\mu$ of
the running coupling $\alpha_s(\mu^2).$ The purpose of the running
coupling in any gauge theory is to sum all terms involving the
$\beta$ function; in fact, when the renormalization scale is set
properly, all non-conformal $\beta \ne 0$ terms  in a perturbative
expansion arising from renormalization are summed into the running
coupling. The remaining terms in the perturbative series are then
identical to that of a conformal theory; i.e., the corresponding
theory with $\beta=0$. The resulting scale-fixed predictions using
the  ``principle of maximum conformality'' (PMC)~\cite{Brodsky:2011ig} are thus independent of
the choice of renormalization scheme --  a key requirement of
renormalization group invariance.   The results avoid renormalon
resummation and agree with QED scale-setting in the Abelian limit.
The PMC is also the theoretical principle underlying the BLM
procedure,~\cite{Brodsky:1982gc} commensurate scale relations~\cite{Lu:1992nt} between observables, and
the scale-setting methods used in lattice gauge theory.  The number
of active flavors $n_f$ in the QCD $\beta$ function is also
correctly determined. 

DiGiustino, Wu,~\cite{Brodsky:2011ig,Brodsky:2011ta} and Brodsky have shown how to determine
the PMC/BLM  scale for QCD processes up to NNLO.~\cite{Brodsky:2011ig}  See also Refs.~ \cite{Grunberg:1991ac,Mikhailov:2004iq,Kataev:2010du}.
A single global
PMC scale, valid at leading order, can be derived from basic
properties of the perturbative QCD cross section. The elimination
of the renormalization scheme ambiguity using the PMC will not
only increase the precision of QCD tests,  but it will also
increase the sensitivity of collider experiments to new physics
beyond the Standard Model.

\section{Vacuum Condensates and the Cosmological Constant}

When one makes a measurement in hadron physics, such as  deep inelastic lepton-proton scattering, one probes  hadron's constituents consistent with causality -- at a given light front time, not at instant time.  Similarly, when one makes observations in cosmology, information is obtained  consistent with the finite speed of light.   

Physical eigenstates are built from operators acting on the vacuum.
It is thus important to distinguish two very different concepts of the vacuum in quantum field theories such as QED and QCD.   The conventional instant-form vacuum is a state defined at the same time $t$ at all spatial points in the universe.  In contrast, the front-form vacuum only senses phenomena which are causally connected; i.e., or within the observer's light-cone.
The instant-form vacuum is defined as the lowest energy eigenstate of the instant-form Hamiltonian.  For example, the instant-form vacuum in QED is saturated with quantum loops of leptons and photons. In calculations of physical processes one must then normal-order the vacuum and divide the $S$-matrix elements by the disconnected vacuum loops.

In contrast, the front-form (light-front) vacuum is defined as the lowest mass eigenstate of light-front Hamiltonian quantized  at fixed $\tau = t -z/c$. The  LF vacuum is remarkably simple  because of the restriction $k^+ \ge 0.$   The LF vacuum thus coincides with the  vacuum of the free LF Hamiltonian  (up to possible zero modes set by boundary conditions).  The front-form vacuum  is causal and Lorentz invariant; whereas the instant form vacuum is acausal and depends on the observer's Lorentz frame. 

In fact, in the LF analysis, one finds that the spatial support of QCD condensates
is restricted to the interior of hadrons, physics which arises due to the
interactions of confined quarks and gluons.  The condensate physics normally associated with the instant-form vacuum is replaced by the dynamics of higher non-valence Fock states as shown in the context of the infinite momentum method by Casher and Susskind.~\cite{Casher:1974xd}  In particular, chiral symmetry is broken in a limited domain of size $1/ m_\pi$,  in analogy to the limited physical extent of superconductor phases.  
This novel description  of chiral symmetry breaking  in terms of ``in-hadron condensates''  has also been observed in Bethe-Salpeter studies.~\cite{Maris:1997hd, Maris:1997tm}
The usual argument for a quark vacuum condensate is the Gell-Mann--Oakes--Renner formula:
$
m^2_\pi = -2 m_q {\langle0| \bar q q |0\rangle/ f^2_\pi}.
$
However, in the Bethe-Salpeter and light-front formalisms, where the pion is a $q \bar q$ bound-state, the GMOR relation is replaced by
$
m^2_\pi = - 2 m_q {\langle 0| \bar q \gamma_5  q |\pi \rangle/ f_\pi},
$
where $\rho_\pi \equiv - \langle0| \bar q \gamma_5  q |\pi\rangle$  represents a pion decay constant via an an elementary pseudoscalar current. 

The cosmological constant measures the matrix element of the energy momentum tensor $T^{\mu \nu}$ in the background universe.  It corresponds to the measurement of the gravitational interactions of  a probe of finite mass;   it only senses the causally connected domain within the light-cone of the observer.  If the universe is empty, the appropriate vacuum state is thus the LF vacuum since it is causal.  One automatically obtains a vanishing cosmological constant from the LF vacuum.
Thus, as argued in Refs. ~\cite{Brodsky:2009zd, Brodsky:2008be, Brodsky:2008xu}   the 45 orders of magnitude conflict of QCD with the observed value of the cosmological condensate is removed, and 
a new perspective on the nature of quark and gluon condensates in
QCD  is thus obtained.

\section{Conclusions \label{conclusions}}

The AdS/QCD correspondence is now providing important insight into the properties of QCD needed to compute exclusive reactions. In particular, the soft-wall AdS/QCD model modified by a positive sign dilaton metric, which represents color confinement, leads to a remarkable one-parameter description of nonperturbative hadron dynamics, including successful predictions for the meson and baryon spectrum for zero quark mass,
including
a zero-mass pion, a Regge spectrum of linear trajectories with the same slope in orbital angular $L$ and the
principal quantum number $n$, as well as dynamical form factors.
The theory predicts
dimensional counting for form factors and other fixed CM angle exclusive reactions.  Moreover, as we have discussed,
light-front holography allows one to map the hadronic amplitudes $\Phi(z)$ determined in the AdS fifth dimension $z$ to the valence LFWFs of each hadron as a function of a covariant impact variable $\zeta.$  Moreover, the same techniques provide a prediction for the QCD coupling $\alpha_s(Q^2)$ and its $\beta$-function which reflects the dynamics of confinement.

The combination of anti-de Sitter  space
methods with light-front  holography provides an accurate first approximation for the spectra and wavefunctions of meson and baryon light-quark  bound states. One also obtains an elegant connection between a semiclassical first approximation to QCD, quantized on the light-front, with hadronic modes propagating on a fixed AdS background. The resulting bound-state Hamiltonian equation of motion in QCD leads to  relativistic light-front wave equations in the invariant impact variable $\zeta$ which measures the separation of the quark and gluonic constituents within the hadron at equal light-front time. This corresponds
to the effective single-variable relativistic Schr\"odinger-like equation in the AdS fifth dimension coordinate $z$,  Eq. (\ref{eq:QCDLFWE}). The eigenvalues give the hadronic spectrum, and the eigenmodes represent the probability distributions of the hadronic constituents at a given scale.
In particular, the soft-wall model  with a positive dilaton profile $\exp(\kappa^2 z^2)$ gives a successful accounting of the spectra of the light-quark meson and baryons~\cite{deTeramond:2009xk}.  

We have also shown that the light-front holographic mapping of  effective classical gravity in AdS space, modified by the positive-sign dilaton background  predicts the form of a non-perturbative effective coupling $\alpha_s^{AdS}(Q)$ and its $\beta$-function.
The AdS/QCD running coupling is in very good agreement with the effective
coupling $\alpha_{g_1}$ extracted from  the Bjorken sum rule.
The holographic $\beta$-function displays a transition from  nonperturbative to perturbative  regimes  at a momentum scale $Q \sim 1$ GeV.
Our analysis indicates that light-front holography captures the characteristics of the full  $\beta$-function of QCD  and  the essential dynamics of confinement, thus giving further support to the application of the gauge/gravity duality to the confining dynamics of strongly coupled QCD.

There are many phenomenological applications where detailed knowledge of the QCD coupling and the renormalized gluon propagator at relatively soft momentum transfer are essential.
This includes exclusive and semi-exclusive processes as well as the rescattering interactions which
create the leading-twist Sivers single-spin correlations in
semi-inclusive deep inelastic scattering,~\cite{Brodsky:2002cx, Collins:2002kn} the Boer-Mulders functions which lead to anomalous  $\cos 2 \phi$  contributions to the lepton pair angular distribution in the unpolarized Drell-Yan reaction,~\cite{Boer:2002ju} and the Sommerfeld-Sakharov-Schwinger correction to heavy-quark production at
threshold.~\cite{Brodsky:1995ds}
The confining  coupling from light-front holography thus can provide a
quantitative understanding of this factorization-breaking physics.~\cite{Collins:2007nk}.

As we have seen, the gauge/gravity duality leads  to a simple analytical and phenomenologically compelling nonperturbative approximation to the full light-front   QCD Hamiltonian. For example, the soft-wall holographic model leads to a remarkable one-parameter description of nonperturbative hadron dynamics despite the fact that explicit gluonic quanta are absent. 
The absence of explicit  gluonic degrees of freedom at small virtuality is a feature of color-confining holographic QCD models which are based on the gauge/gravity duality  on warped five-dimensional anti-de Sitter spaces following the original Maldacena correspondence.  The sublimation of soft gluons can explain several outstanding anomalies of hadron physics including the absence of the Landshoff mechanism in large-angle hadron-hadron scattering and the $J/\psi \to \rho \pi $ puzzle.

\section*{Acknowledgments}

Invited talk presented by SJB at the International Workshop on QCD Green's Functions, Confinement and Phenomenology,
5-9 September 2011, Trento, Italy.
We thank  Leonardo di Giustino, Xing-Gang Wu, Mike Cornwall, Sadataka Furui, 
Craig Roberts, Robert Shrock, and Peter Tandy for helpful comments.
We are  also grateful to  A. Deur, H. G. Dosch, J. Erlich,  and F.-G. Cao for their collaborations. This research was supported by the Department
of Energy  contract DE--AC02--76SF00515. SLAC-PUB--14843
.

  \end{document}